\documentclass[11pt, draftclsnofoot, onecolumn]{IEEEtran}

\ifCLASSINFOpdf
\else
\fi
\usepackage{amsmath,amssymb,latexsym,cite}
\usepackage{epsf, pgf, xcolor}
\usepackage{pstricks}
\usepackage{enumerate}
\usepackage{algorithm, algorithmic}
\usepackage{booktabs, ctable}
\usepackage{subfigure}
\usepackage{amsfonts}
\usepackage{bm}
\usepackage{epsfig}
\usepackage{array, tabularx}
\usepackage{pst-node}
\usepackage{cases}
\usepackage{etex}
\usepackage{pmat}
\usepackage{pst-sigsys}
\usepackage[top=3.4cm, bottom=2.35cm, left=2.2cm, right=2.3cm]{geometry}

\newcommand{\mng}[1]{\mathbf{#1}}

\newrgbcolor{Sgray}{0.981 0.981 0.981}
\newrgbcolor{Rgray}{0.94 0.94 0.94}
\newrgbcolor{Dgray}{0.8 0.8 0.8}

\def\EE{E}
\def\TR{\mbox{tr}}

\def\LTN{LTN}
\def\CC{\mathcal{C}}
\def\VV{\mathcal{V}}
\def\EDG{\mathcal{E}}

\def\os2{\sigma_S^2}
\def\on2{\sigma_N^2}
\def\oz2{\sigma_Z^2}

\newtheorem{theorem}{Theorem}

\newtheorem{lemma}{Lemma}

\newtheorem{definition}{Definition}
\newtheorem{example}{Example}
\newtheorem{remark}{Remark}

\IEEEoverridecommandlockouts

\begin{document}
%

\title{Reduced-Dimension Linear Transform Coding of Correlated Signals in Networks}


%
%
%

\author{Naveen~Goela,~\IEEEmembership{Student Member,~IEEE,} 
        and~Michael~Gastpar$^\dagger$,~\IEEEmembership{Member,~IEEE}
\thanks{This work was supported in part by the National Science Foundation under Grant CCF-0627024 and made with U.S. Government support under and awarded by DoD, Air Force Office of Scientific Research, National Defense Science and Engineering Graduate (NDSEG) Fellowship, 32 CFR 168a. The material in this paper was presented in part at the IEEE International Conference on Acoustics, Speech and Signal Processing, Taipei, Taiwan in April 2009 and at the IEEE International Symposium on Information Theory, Seoul, South Korea in June 2009.}
\thanks{Copyright~\copyright\,2012 IEEE. Personal use of this material is permitted. However,
permission to use this material for any other purposes must be obtained from the
IEEE by sending a request to pubs-permissions@ieee.org.}
\thanks{N. Goela and M. Gastpar are with the Department of Electrical Engineering and Computer Science, University of California, Berkeley, Berkeley, CA 94720-1770 USA (e-mail: \{ngoela, gastpar\}@eecs.berkeley.edu).}
\thanks{$^\dagger$M. Gastpar is also with the School of Computer and Communication Sciences, Ecole Polytechnique F\'ed\'erale (EPFL), Lausanne, Switzerland.}
}

%
%


\markboth{To Appear in IEEE Transactions on Signal Processing}%
{Goela and Gastpar: Reduced-Dimension Linear Transform Coding of Correlated Signals in Networks}

%



\maketitle

\begin{abstract}
A model, called the linear transform network (\LTN), is proposed to analyze the compression and estimation of correlated signals transmitted over directed acyclic graphs (DAGs). An \LTN~is a DAG network with multiple source and receiver nodes. Source nodes transmit subspace projections of random correlated signals by applying reduced-dimension linear transforms. The subspace projections are linearly processed by multiple relays and routed to intended receivers. Each receiver applies a linear estimator to approximate a subset of the sources with minimum mean squared error (MSE) distortion. The model is extended to include noisy networks with power constraints on transmitters. A key task is to compute all local compression matrices and linear estimators in the network to minimize end-to-end distortion. The non-convex problem is solved iteratively within an optimization framework using constrained quadratic programs (QPs). The proposed algorithm recovers as special cases the regular and distributed Karhunen-Lo\`eve transforms (KLTs). Cut-set lower bounds on the distortion region of multi-source, multi-receiver networks are given for linear coding based on convex relaxations. Cut-set lower bounds are also given for any coding strategy based on information theory. The distortion region and compression-estimation tradeoffs are illustrated for different communication demands (e.g. multiple unicast), and graph structures. 

\end{abstract}


%
\IEEEpeerreviewmaketitle

\section{Introduction}
%
%
%
%
\IEEEPARstart{T}{he} compression and estimation of an observed signal via subspace projections is both a classical and current topic in signal processing and communication. While random subspace
projections have received considerable attention in the compressed sensing literature~\cite{donoho06}, subspace projections optimized for minimal distortion are important for many applications. The Karhunen-Lo\`eve transform (KLT) and its empirical form Principal Components Analysis (PCA), are widely studied in computer vision, biology, signal processing, and information theory. Reduced dimensionality representations are useful for source coding,
noise filtering, compression, clustering, and data mining. Specific examples include eigenfaces for face recognition, orthogonal decomposition in transform coding, and sparse PCA for gene analysis~\cite{turkpentland91, goyal01, ghaouijordan07}.

In contemporary applications such as wireless sensor networks (WSNs)
and distributed databases, data is available and collected in
different locations. In a WSN, sensors are usually constrained by
limited power and bandwidth resources. This has motivated existing
approaches to take into account correlations across high-dimensional
sensor data to reduce transmission requirements (see e.g.~\cite{gastpar06, giannakis07, xiao08, zhangphd, fang10, zhu05, fang_and_li08}). Rather than transmitting raw sensor data to a fusion center to approximate a global signal, sensor nodes carry out local data dimensionality reduction to increase bandwidth and energy efficiency.


In the present paper, we propose a linear transform network (\LTN) model to analyze dimensionality reduction for compression-estimation of correlated signals in \emph{multi-hop} networks. In a centralized setting, given a random source signal $\pmb{x}$ with zero-mean and covariance matrix $\mng{\Sigma}_{\pmb{x}}$, applying the KLT to $\pmb{x}$ yields uncorrelated components in the eigenvector basis of $\mng{\Sigma}_{\pmb{x}}$. The optimal linear least squares $k^{th}$-order approximation of the source is given by the $k$ components corresponding to the $k$ largest eigenvalues of $\mng{\Sigma}_{\pmb{x}}$. In a network setting, multiple correlated signals are observed by different source nodes. The source nodes transmit low-dimensional subspace projections (approximations of the source) to intended receivers via a relay network. The compression-estimation problem is to optimize the subspace projections computed by all nodes in order to minimize the end-to-end distortion at receiver nodes.

In our model, receivers estimate random vectors based on ``one-shot'' linear \emph{analog-amplitude} multisensor observations. The restriction to ``one-shot'', zero-delay encoding of each vector of source observations separately is interesting due to severe complexity limitations in many applications (e.g. sensor networks). Linear coding depends on first-order and second-order statistics and is robust to uncertainty in the precise probabilistic distribution of the sources. Under the assumption of ideal channels between nodes, our task is to optimize signal subspaces given limited bandwidth in terms of the number of real-valued messages communicated. Our results extend previous work on distributed estimation in this case~\cite{gastpar06, giannakis07, xiao08, zhangphd}. For the case of dimensionality-reduction with noisy channel communication (see e.g.~\cite{giannakis07}), the task is to optimize signal subspaces subject to channel noise and power constraints.

For noisy networks, the general communication problem is often referred to as the \emph{joint source-channel-network coding problem} in the information-theoretic literature and is a famously open problem. Beyond the zero-delay, linear dimensionality-reduction considered here, end-to-end performance in networks could be improved by (i), non-linear strategies and (ii), allowing a longer coding horizon. Partial progress includes non-linear low-delay mappings for only simple network scenarios~\cite{ramstad02,skoglund06,hekland09}. For the case of an infinite coding horizon, separation theorems for decomposing the joint communication problem have been analyzed by~\cite{ramamoorthy06,jalali10,Gastpar03dimacs}.

\subsection{Related Work}
\label{sec:RelatedWork}
Directly related to our work in networks is the \emph{distributed} KLT problem. Distributed linear transforms were introduced by Gastpar et al. for the compression of jointly Gaussian sources using iterative methods~\cite{gastpar06}\cite{gastpar02}. Simultaneous work by Zhang et al. for multi-sensor data fusion also resulted in iterative procedures~\cite{zhangphd}. An alternate proof based on innovations
for second order random variables with arbitrary distributions was
given by~\cite{nurdin09}. The problem was extended for non-Gaussian
sources, including channel fading and noise effects to model the
non-ideal link from sensors to decoder by Schizas et al.~\cite{giannakis07}\nocite{xiao08}. Roy and Vetterli provide an \emph{asymptotic} distortion analysis of the distributed KLT, in the case when the dimension of the source and observation vectors approaches infinity~\cite{royvetterli08}.
Finally, Xiao et al. analyze linear transforms for distributed
\emph{coherent} estimation~\cite{xiao08}.

Much of the estimation-theoretic literature deals with
\emph{single-hop} networks; each sensor relays information directly
to a fusion center. In \emph{multi-hop} networks, linear operations
are performed by successive relays to aggregate, compress, and
redistribute correlated signals. The \LTN~model relates to recent work on routing and \emph{network coding}~(Ahlswede et al.~\cite{acly00}). In pure routing solutions, intermediate nodes either forward or drop packets. The corresponding analogy in the \LTN~model is to constrain transforms to be essentially identity transforms. However, network coding (over finite fields) has shown that mixing of data at intermediate nodes achieves higher rates in the multicast setting (see~\cite{lyc03} regarding the sufficiency of linear codes and~\cite{jaggi05} for multicast code construction). Similarly in the \LTN~model, linear combining of subspace projections (over the real field) at intermediate nodes improves decoding performance. Lastly, the max-flow min-cut theorem of Ford-Fulkerson~\cite{fulkerson56} provides the basis for cut-set lower bounds in networks.

The \LTN~model is partially related to the formulation of Koetter and Kschischang~\cite{koetter08} modeling information transmission as the injection of a basis for a vector space into the
network, and subspace codes~\cite{silva08}. If arbitrary data exchange is permitted between network nodes, the compression-estimation problem is related to estimation in graphical models (e.g. decomposable PCA~\cite{wiesel08}, and tree-based transforms (tree-KLT)~\cite{ortega09}). Other related work involving signal projections in networks includes joint source-channel communication in sensor networks~\cite{nowak07}, random projections in a gossip framework\cite{rabbat06}, and distributed compressed sensing~\cite{baraniuk05}.

\subsection{Summary of Main Results}
\label{sec:SummaryMainResults}
We cast the network compression-estimation problem as a statistical signal processing and constrained optimization problem. For most networks, the optimization is non-convex. Therefore, our main results are divided into two categories: (i) Iterative solutions for linear transform coding over acyclic networks; (ii) Cut-set bounds based on convex relaxations and cut-set bounds based on information theory.
\begin{itemize}
\item Section~\ref{sec:LinearProcessingNetworks} reviews linear signal processing in networks. Section~\ref{sec:IterativeMethod} outlines an iterative optimization for compression-estimation matrices in ideal networks under a local convergence criterion.
\item Section~\ref{sec:NoisyModel} analyzes an iterative optimization method involving constrained quadratic programs for noisy networks with power allocation over subspaces.
\item Section~\ref{sec:CutsetBounds} introduces cut-set lower bounds to benchmark the minimum mean square error (MSE) for linear coding based on convex relaxations such as a semi-definite program (SDP) relaxation. 
\item Section~\ref{sec:AppendixCutSetBoundsInfoTheory} describes cut-set lower bounds for any coding strategy in networks based on information-theoretic principles of source-channel separation. The lower bounds are plotted for a distributed noisy network. 
\item Sections~\ref{sec:IterativeMethod}-\ref{sec:CutsetBounds} provide examples illustrating the tradeoffs between compression and estimation; upper and lower bounds are illustrated for an aggregation (tree) network, butterfly network, and distributed noisy network.
\end{itemize}




\subsection{Notation}
\label{sec:Notation}


Boldface upper case letters denote matrices, boldface lower case letters denote column
vectors, and calligraphic upper case letters denote sets. The $\ell^2$-norm of a vector $\pmb{x} \in \mathbb{R}^{n}$ is defined as $\|\pmb{x}\|_{2} \triangleq \sqrt{\sum_{i=1}^{n} |x_i|^2}$. The
weighted $\ell^2$-norm $\left\|\pmb{x}\right\|_{\mng{W}} \triangleq \left\|\mng{W}\pmb{x}\right\|_{2}$ where $\mng{W}$ is a positive semi-definite matrix (written $\mng{W} \succeq \mng{0}$). Let $(\cdot)^{T}$, $(\cdot)^{-1}$, and $\TR(\cdot)$ denote matrix transpose, inverse, and trace respectively. Let $\mng{A} \otimes \mng{B}$ denote the Kronecker matrix product of two matrices. The matrix $\mng{I}_{\ell}$ denotes the $\ell \times \ell$ identity. For $\ell \geq k$, the notation $\mng{T}_{k:\ell} \triangleq \mng{T}_k \mng{T}_{k+1} \cdot \cdot \cdot \mng{T}_{\ell}$ denotes the product of $(\ell - k + 1)$ matrices. A matrix $\mng{X} \in \mathbb{R}^{m \times n}$ is written in vector form $\mbox{vec}(\mng{X}) \in \mathbb{R}^{mn}$ by stacking its columns; i.e. $\mbox{vec}(\mng{X}) = [\pmb{x}_1;\pmb{x}_2;~\ldots~;\pmb{x}_n]$ where $\pmb{x}_j$ is the $j$-th column of $\mng{X}$. For random vectors, $\EE[\cdot]$ denotes the expectation, and $\mng{\Sigma}_{\pmb{x}} \triangleq \EE[\pmb{x}\pmb{x}^{T}]$ denotes the covariance matrix of the zero-mean random vector $\pmb{x}$.

\section{Problem Statement}\label{sec:ProblemStatement}
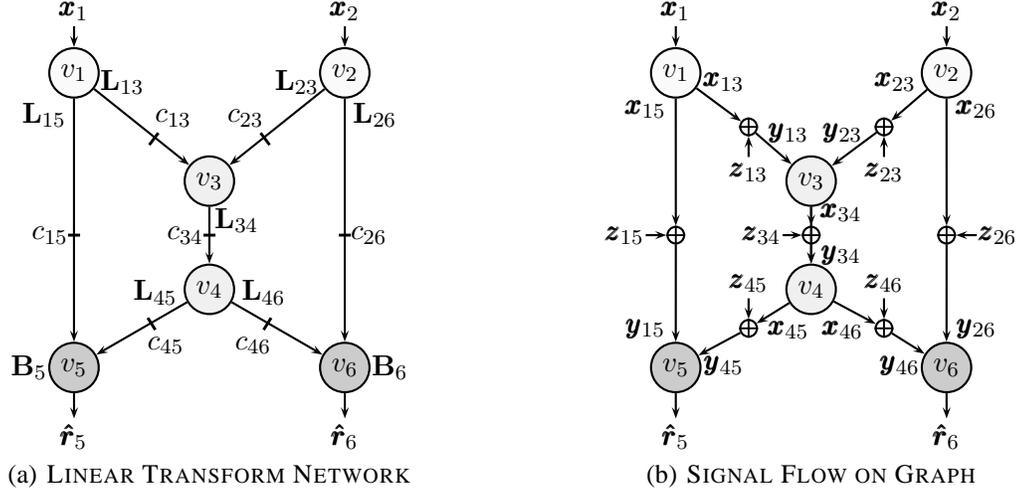
\begin{figure*}[t]
\begin{center}
\psset{unit=0.40mm}
\begin{pspicture}(-11,-162)(115,-3)

\rput(-100,0){


\rput(-32,-2){


\rput(-5,0){
\cnodeput[linestyle=solid,linecolor=black,fillcolor=Sgray,fillstyle=solid](51,-31){A}{$v_1$}
\cnodeput[linestyle=solid,linecolor=black,fillcolor=Dgray,fillstyle=solid](51,-129){B}{$v_5$}
\ncline{->}{A}{B}

\pnode(51,-15.5){S1} \ncline{->}{S1}{A}
\rput(51,-10.5){$\pmb{x}_{1}$}

\pnode(51,-146){T1} \ncline{->}{B}{T1}

\rput(51,-152){$\pmb{\hat{r}}_5$}

\rput(41,-45){$\mng{L}_{15}$} \rput(36, -129){$\mng{B}_{5}$}
\rput(67,-34){$\mng{L}_{13}$}

}

\rput(5,0){
\cnodeput[linestyle=solid,linecolor=black,fillcolor=Sgray,fillstyle=solid](131,-31){C}{$v_2$}
\cnodeput[linestyle=solid,linecolor=black,fillcolor=Dgray,fillstyle=solid](131,-129){D}{$v_6$}
\ncline{->}{C}{D}

\pnode(131,-15.5){S2} \ncline{->}{S2}{C}
\rput(131,-10.5){$\pmb{x}_{2}$}

\pnode(131,-146){T2} \ncline{->}{D}{T2}

\rput(131,-152){$\pmb{\hat{r}}_6$}

\rput(141,-45){$\mng{L}_{26}$} \rput(146, -129){$\mng{B}_{6}$}
\rput(115,-34){$\mng{L}_{23}$}

}

\cnodeput[linestyle=solid,linecolor=black,fillcolor=Rgray,fillstyle=solid](91,-67){E}{$v_3$}
\cnodeput[linestyle=solid,linecolor=black,fillcolor=Rgray,fillstyle=solid](91,-103){F}{$v_4$}
\ncline{->}{E}{F} \ncline{->}{A}{E} \ncline{->}{C}{E}
\ncline{->}{F}{B} \ncline{->}{F}{D}

\rput(100,-80){$\mng{L}_{34}$} \rput(109,-104){$\mng{L}_{46}$}
\rput(73,-104){$\mng{L}_{45}$}

\rput(91, -165){\small{(a) \textsc{Linear Transform Network}}}

\pnode(89,-85){M1} \pnode(93,-85){M2}
\ncline[linewidth=1.1pt]{-}{M1}{M2} \rput(83,-85){$c_{34}$} 

\pnode(44,-85){M3} \pnode(48,-85){M4}
\ncline[linewidth=1.1pt]{-}{M3}{M4} \rput(38,-85){$c_{15}$} 

\pnode(134,-85){M5} \pnode(138,-85){M6}
\ncline[linewidth=1.1pt]{-}{M5}{M6} \rput(144,-85){$c_{26}$} 

\pnode(71,-55){M7} \pnode(74,-51){M8}
\ncline[linewidth=1.1pt]{-}{M7}{M8} \rput(79,-46){$c_{13}$} 

\pnode(111,-55){M9} \pnode(108,-51){M10}
\ncline[linewidth=1.1pt]{-}{M9}{M10} \rput(103,-46){$c_{23}$} 

\pnode(111.5,-112.5){M11} \pnode(109,-116.5){M12}
\ncline[linewidth=1.1pt]{-}{M11}{M12} \rput(105.5,-121.5){$c_{46}$} 

\pnode(70.5,-112.5){M13} \pnode(73,-116.5){M14}
\ncline[linewidth=1.1pt]{-}{M13}{M14} \rput(76.5,-121.5){$c_{45}$} 

}

}

\rput(100,0){


\rput(-32,-2){


\rput(-5,0){
\cnodeput[linestyle=solid,linecolor=black,fillcolor=Sgray,fillstyle=solid](51,-31){A}{$v_1$}
\cnodeput[linestyle=solid,linecolor=black,fillcolor=Dgray,fillstyle=solid](51,-129){B}{$v_5$}

\pnode(51,-15.5){S1} \ncline{->}{S1}{A}
\rput(51,-10.5){$\pmb{x}_{1}$}

\pnode(51,-146){T1} \ncline{->}{B}{T1}

\rput(51,-152){$\pmb{\hat{r}}_{5}$}

\rput(41,-116){$\pmb{y}_{15}$}

\rput(67, -128.5){$\pmb{y}_{45}$}

\rput(41,-43){$\pmb{x}_{15}$}

\rput(67,-34){$\pmb{x}_{13}$}

\pscircleop[opsep=0](51,-85){op15}
\pnode(51,-81){Z15VarA}
\pnode(51,-89){Z15VarB}
\pnode(47,-85){Z15VarC}
\pnode(41,-85){Z15VarD}
\ncline{->}{Z15VarD}{op15}
\rput(34,-85){$\pmb{z}_{15}$}
\ncline{->}{A}{op15} \ncline{->}{op15}{B}

}

\rput(5,0){
\cnodeput[linestyle=solid,linecolor=black,fillcolor=Sgray,fillstyle=solid](131,-31){C}{$v_2$}
\cnodeput[linestyle=solid,linecolor=black,fillcolor=Dgray,fillstyle=solid](131,-129){D}{$v_6$}

\pnode(131,-15.5){S2} \ncline{->}{S2}{C}
\rput(131,-10.5){$\pmb{x}_{2}$}

\pnode(131,-146){T2} \ncline{->}{D}{T2}

\rput(131,-152){$\pmb{\hat{r}}_{6}$}

\rput(141, -116){$\pmb{y}_{26}$}

\rput(115.5, -128.5){$\pmb{y}_{46}$}

\rput(141,-43){$\pmb{x}_{26}$}

\rput(114,-34){$\pmb{x}_{23}$}

\pscircleop[opsep=0](131,-85){op26}
\pnode(131,-81){Z26VarA}
\pnode(131,-89){Z26VarB}
\pnode(135,-85){Z26VarC}
\pnode(141,-85){Z26VarD}
\ncline{->}{Z26VarD}{op26}
\rput(148,-85){$\pmb{z}_{26}$}
\ncline{->}{C}{op26} \ncline{->}{op26}{D}

}

\cnodeput[linestyle=solid,linecolor=black,fillcolor=Rgray,fillstyle=solid](91,-67){E}{$v_3$}
\cnodeput[linestyle=solid,linecolor=black,fillcolor=Rgray,fillstyle=solid](91,-103){F}{$v_4$}
\ncline{->}{E}{F}

\rput(83.5,-116.5){$\pmb{x}_{45}$}

\rput(101.5,-116.5){$\pmb{x}_{46}$}

\rput(101,-91.5){$\pmb{y}_{34}$}

\rput(101,-77.5){$\pmb{x}_{34}$}

\rput(83.5,-51){$\pmb{y}_{13}$}

\rput(101.5,-51){$\pmb{y}_{23}$}

\rput(91, -165){\small{(b) \textsc{Signal Flow on Graph}}}

\pscircleop[opsep=0](91,-85){op34}
\pnode(91,-81.5){Z34VarA}
\pnode(91,-88.5){Z34VarB}
\pnode(87.5,-85){Z34VarC}
\pnode(81.5,-85){Z34VarD}
\ncline{->}{Z34VarD}{op34}
\rput(74.5,-85){$\pmb{z}_{34}$}
\ncline{->}{E}{op34} \ncline{->}{op34}{F}

\rput(-62.5, 36){
\pscircleop[opsep=0](131,-85){op13}
\pnode(128.1716,-82.1716){Z13VarA}
\pnode(133.8284,-87.8284){Z13VarB}
\pnode(131,-89){Z13VarC}
\pnode(131,-95){Z13VarD}
\ncline{->}{Z13VarD}{op13}
\rput(131,-100){$\pmb{z}_{13}$}
}
\rput(-62.5, -31){
\pscircleop[opsep=0](131,-85){op45}
\pnode(134.3538,-82.8200){Z45VarA}
\pnode(127.6462,-87.1800){Z45VarB}
\pnode(131,-81){Z45VarC}
\pnode(131,-75){Z45VarD}
\ncline{->}{Z45VarD}{op45}
\rput(131,-70){$\pmb{z}_{45}$}
}
\rput(-17.5, 36){
\pscircleop[opsep=0](131,-85){op23}
\pnode(133.8284,-82.1716){Z23VarA}
\pnode(128.1716,-87.8284){Z23VarB}
\pnode(131,-89){Z23VarC}
\pnode(131,-95){Z23VarD}
\ncline{->}{Z23VarD}{op23}
\rput(131,-100){$\pmb{z}_{23}$}
}
\rput(-17.5, -31){
\pscircleop[opsep=0](131,-85){op46}
\pnode(127.6462,-82.8200){Z46VarA}
\pnode(134.3538,-87.1800){Z46VarB}
\pnode(131,-81){Z46VarC}
\pnode(131,-75){Z46VarD}
\ncline{->}{Z46VarD}{op46}
\rput(131,-70){$\pmb{z}_{46}$}
}
\ncline{->}{A}{op13} \ncline{->}{op13}{E}
\ncline{->}{C}{op23} \ncline{->}{op23}{E}
\ncline{->}{F}{op45} \ncline{->}{op45}{B}
\ncline{->}{F}{op46} \ncline{->}{op46}{D}

}

}

\end{pspicture}
\end{center}
\vspace{0.1in} \caption{(a) \emph{Linear Transform Network:} An
\LTN~model with source nodes $\{v_1, v_2\}$ and receivers $\{v_5, v_6\}$. Source nodes observe vector signals $\{\pmb{x}_1, \pmb{x}_2\}$. All encoding nodes linearly process received signals using a transform $\mng{L}_{ij}$. Receivers $v_5$ and $v_6$ compute LLSE estimates $\pmb{\hat{r}}_{5}$ and $\pmb{\hat{r}}_{6}$ of desired signals $\pmb{r}_{5}$ and $\pmb{r}_6$. (b) \emph{Signal Flow Graph:} Linear processing of source signals $\{\pmb{x}_1, \pmb{x}_2\}$ results in signals transmitted along edges of the graph.}\vspace{-0.1in} \label{fig:PosterNetwork}
\end{figure*}

Fig.~\ref{fig:PosterNetwork} serves as an extended example of an
\LTN~graph. The network is comprised of two sources, two relays, and
two receiver nodes.
\begin{definition}[Relay Network]\label{def:Graph} Consider
a relay network modeled by a directed acyclic graph (DAG) $G = (\VV, \EDG)$ and a set of
weights $\CC$. The set $\VV = \{v_1, v_2, \ldots, v_{|\VV|}\}$ is
the vertex/node set, $\EDG \subset \{1,\ldots,|\VV|\} \times
\{1,\ldots,|\VV|\}$ is the edge set, and $\CC = \{c_{ij} \in
\mathbb{Z}^{+} : (i,j) \in \EDG\}$ is the set of weights. Each edge
$(i,j) \in \EDG$ represents a communication link with integer
bandwidth $c_{ij}$ from node $v_i$ to $v_j$. The in-degree and out-degree of a
node $v_i$ are computed as
\begin{align}
d^{-}_i & = \sum_{q:(q,i) \in \EDG} c_{qi}, \\
d^{+}_i & = \sum_{l:(i,l) \in \EDG} c_{il}.
\end{align}
\end{definition}
As an example, the graph in Fig.~\ref{fig:PosterNetwork} consists of
nodes $\VV = \{v_1, v_2, \ldots, v_6\}$. Integer bandwidths
$c_{ij}$ for each communication link $(i,j)$ are marked.


\begin{definition}[Source and Receiver Nodes]\label{def:SourceDestNodes}
Given a relay network $G = (\VV,\EDG)$, the set of source nodes
$\mathcal{S} \subset \VV$ is defined as $\mathcal{S} = \{v_i \in \VV
~|~ d^{-}_i = 0\}$. We assume a labeling of nodes in $\VV$ so that
$\mathcal{S} = \{v_1, v_2, \ldots, v_{|\mathcal{S}|}\}$, i.e. the
first $|\mathcal{S}|$ nodes are source nodes. The set of receiver
nodes $\mathcal{T} \subset \VV$ is defined as $\mathcal{T} = \{v_i
\in \VV ~|~ d^{+}_i = 0\}$.\footnote{For networks of interest in this
paper, an arbitrary DAG $G$ may be augmented with auxiliary nodes to
ensure that source nodes have in-degree $d^{-}_i = 0$ and
receiver nodes have out-degree $d^{+}_i = 0$.} Let $\kappa
\triangleq |\VV| - |\mathcal{T}|$. We assume a labeling of nodes in
$\VV$ so that $\mathcal{T} = \{v_{\kappa + 1}, v_{\kappa + 2},
\ldots, v_{|\VV|}\}$, i.e. the last $|\mathcal{T}|$ nodes are
receiver nodes.
\end{definition}
In Fig.~\ref{fig:PosterNetwork}, $\mathcal{S} = \{v_1, v_2\}$ and
$\mathcal{T} = \{v_5, v_6\}$. 
\subsection{Source Model}
\begin{definition}[Basic Source Model]\label{def:BasicSourceModel} Given a relay
network $G = (\VV,\EDG)$ with source/receiver nodes
$(\mathcal{S}, \mathcal{T})$, the source nodes $\mathcal{S} = \{v_i\}_{i=1}^{|\mathcal{S}|}$ observe random signals $\mathcal{X} = \{\pmb{x}_i\}_{i=1}^{|\mathcal{S}|}$. The random vectors $\pmb{x}_i \in \mathbb{R}^{n_i}$ are assumed zero-mean with covariance $\mng{\Sigma}_{ii}$, and cross-covariances $\mng{\Sigma}_{ij} \in
\mathbb{R}^{n_i \times n_j}$. Let $n \triangleq \sum_i n_i$. The distributed network sources may be grouped into an $n$-dimensional random vector $\pmb{x} = [\pmb{x}_1;\pmb{x}_2;~\ldots~;\pmb{x}_{|\mathcal{S}|}]$
with known second-order statistics $\mng{\Sigma}_{\pmb{x}} \in \mathbb{R}^{n \times n}$,
\begin{align}
\mng{\Sigma}_{\pmb{x}} & = \left[\begin{array}{cccc}
\mng{\Sigma}_{11}
& \mng{\Sigma}_{12} & \ldots & \mng{\Sigma}_{1|\mathcal{S}|} \\
\mng{\Sigma}_{21} &
\mng{\Sigma}_{22} & \ldots & \mng{\Sigma}_{2|\mathcal{S}|} \\
\vdots & \vdots & \ddots & \vdots \\ \mng{\Sigma}_{|\mathcal{S}|1} &
\mng{\Sigma}_{|\mathcal{S}|2} & \ldots &
\mng{\Sigma}_{|\mathcal{S}||\mathcal{S}|} \end{array}\right].
\end{align}
\end{definition}
More generally, each source node $v_i \in \mathcal{S}$ emits independent and identically distributed ($i.i.d.$) source vectors $\{\pmb{x}_i[t]\}_{t>0}$ for $t$ a discrete time index; however, in the analysis of zero-delay linear coding, we do not write the time indices explicitly.
\begin{remark} A common linear signal-plus-noise model for sensor networks is of the form $\pmb{x}_{i} = \mng{H}_{i}\pmb{x} + \pmb{n}_{i}$; however, neither a linear source model nor the specific distribution of $\pmb{x}_i$ is assumed here. \emph{A} \emph{priori} knowledge of second-order statistics may be obtained during a training phase via sample estimation.
\end{remark}
In Fig.~\ref{fig:PosterNetwork}, two source nodes $\mathcal{S} =
\{v_1, v_2\}$ observe the corresponding random signals in
$\mathcal{X} = \{\pmb{x}_1, \pmb{x}_2\}$.

\subsection{Communication Model}

\begin{definition}[Communication Model]\label{def:CommModel} Given a
relay network $G = (\VV,\EDG)$ with weight-set $\CC$, each edge $(i,j) \in \EDG$ represents a communication link of bandwidth $c_{ij}$ from $v_i$ to $v_j$. The bandwidth is the dimension of the vector channel. We denote signals exiting $v_i \in \VV$ along edge $(i,j) \in \EDG$ by $\pmb{x}_{ij} \in \mathbb{R}^{c_{ij}}$ and signals entering node $v_j$ along edge $(i,j) \in \EDG$ by $\pmb{y}_{ij} \in \mathbb{R}^{c_{ij}}$. If communication is noiseless, $\pmb{y}_{ij} = \pmb{x}_{ij}$. For all relay nodes and receiver nodes, we further define $\pmb{y}_j \in \mathbb{R}^{d_j^{-}}$ to be the concatenation of all signals $\pmb{y}_{ij}$ incident to node $v_j$ along edges $(i,j) \in \EDG$.

A noisy communication link $(i,j) \in \EDG$ is modeled as: $\pmb{y}_{ij} = \pmb{x}_{ij} + \pmb{z}_{ij}$. The channel
noise $\pmb{z}_{ij} \in \mathbb{R}^{c_{ij}}$ is a Gaussian random vector with zero-mean and covariance $\mng{\Sigma}_{\pmb{z}_{ij}}$. The channel input is power constrained so that $\EE[\|\pmb{x}_{ij}\|_{2}^{2}] \leq P_{ij}$. The power constraints for a network are given by set $\mathcal{P} = \{P_{ij} \in \mathbb{R}^{+} :
(i,j) \in \EDG\}$. The signal-to-noise ratio (SNR) along a link is
\begin{align}
SNR_{ij} & = \frac{\EE\left[\left\| \pmb{x}_{ij}
\right\|^2_2\right]}{\EE\left[\left\| \pmb{z}_{ij} \right\|^2_2\right]}.
\end{align}
\end{definition}
Fig.~\ref{fig:PosterNetwork}(b) illustrates the signal flow of an \LTN~graph.

\subsection{Linear Encoding over Graph $G$}
Source and relay nodes encode random vector signals by applying reduced-dimension linear transforms.
\begin{definition}[Linear Encoding]\label{def:EncodingModel} Given a relay network $G = (\VV,\EDG)$, weight-set $\CC$, source/receiver nodes $(\mathcal{S}, \mathcal{T})$, sources $\mathcal{X}$, and the communication model of Definition~\ref{def:CommModel}, the linear encoding matrices for $G$ are denoted by set $\mathcal{L}_{G} = \{\mng{L}_{ij} : (i,j) \in \EDG\}$. Each $\mng{L}_{ij}$ represents the linear transform applied by node $v_i$ in communication with node $v_j$. For $v_i \in \mathcal{S}$, transform $\mng{L}_{ij}$ is of size $c_{ij} \times n_i$ and represents the encoding $\pmb{x}_{ij} = \mng{L}_{ij}\pmb{x}_i$. For a relay $v_i$, transform $\mng{L}_{ij}$ is of size $c_{ij} \times d^{-}_i$, and $\pmb{x}_{ij} = \mng{L}_{ij}\pmb{y}_i$. The \emph{compression ratio} along edge $(i,j) \in \EDG$ is
\begin{subnumcases}{\alpha_{ij} = }
\frac{c_{ij}}{n_i} & $\mbox{if}~v_i \in \mathcal{S}$, \\
\frac{c_{ij}}{d^{-}_i} & $\mbox{if}~v_i \in \VV \setminus \mathcal{S}$.
\end{subnumcases}
\end{definition}
In Fig.~\ref{fig:PosterNetwork}, the linear encoding matrices for
source node $v_1$ and $v_2$ are $\{\mng{L}_{15}, \mng{L}_{13}\}$ and
$\{\mng{L}_{26}, \mng{L}_{23}\}$ respectively. The linear encoding
matrices for the relays are $\mng{L}_{34}$, $\mng{L}_{45}$,
$\mng{L}_{46}$. The output signals of source
node $v_1$ are $\pmb{x}_{15} = \mng{L}_{15}\pmb{x}_1$ and
$\pmb{x}_{13} = \mng{L}_{13}\pmb{x}_1$. Similarly, the output signal of relay $v_3$ is
\begin{align}
\pmb{x}_{34} = \mng{L}_{34}\pmb{y}_{3} =
\mng{L}_{34}\left[\begin{array}{c} \pmb{y}_{13} \\
\pmb{y}_{23} \end{array}\right].
\end{align}

\subsection{Linear Estimation over $G$}

\begin{definition}[Linear Estimation]\label{def:DecodingModel}
Given relay network $G = (\VV,\EDG)$, weight-set $\CC$, source/receiver
nodes $(\mathcal{S}, \mathcal{T})$, sources $\mathcal{X}$, and the communication model of Def.~\ref{def:CommModel}, the set of linear decoding matrices is denoted $\mathcal{B}_{G} = \{\mng{B}_i\}_{i:v_i \in \mathcal{T}}$. Each receiver $v_i \in \mathcal{T}$ estimates a (zero-mean) random vector $\pmb{r}_i \in \mathbb{R}^{r_i}$ which is correlated with the sources in $\mathcal{X}$. We assume that the second-order statistics $\mng{\Sigma}_{\pmb{r}_i}$, $\mng{\Sigma}_{\pmb{r}_i \pmb{x}}$ are known. Receiver $v_i \in \mathcal{T}$ applies a linear estimator given by matrix $\mng{B}_i \in \mathbb{R}^{r_i \times d^{-}_i}$ to estimate $\pmb{r}_i$ given its observations and computes $\pmb{\hat{r}}_i = \mng{B}_i\pmb{y}_i$. The linear least squares estimate (LLSE) of $\pmb{r}_i$ is denoted by $\pmb{\hat{r}}_i$.
\end{definition}
\vspace{-0.05in}
In Fig.~\ref{fig:PosterNetwork}, receiver $v_5$ reconstructs $\pmb{r}_5$ while receiver $v_6$ reconstructs $\pmb{r}_6$. The LLSE signals $\pmb{\hat{r}}_5$ and $\pmb{\hat{r}}_6$ are computed as
\begin{align}
\pmb{\hat{r}}_5 & = \mng{B}_5 \pmb{y}_5 = \mng{B}_5 \left[\begin{array}{c} \pmb{y}_{15} \\ \pmb{y}_{45} \end{array}\right], \\
\pmb{\hat{r}}_6 & = \mng{B}_6 \pmb{y}_6 = \mng{B}_6
\left[\begin{array}{c} \pmb{y}_{26} \\ \pmb{y}_{46}
\end{array}\right].
\end{align}

\begin{definition}[Distortion Metric]\label{def:DistMetric}
Let $\pmb{x}$ and $\pmb{y}$ be two real vectors of the same dimension. The MSE distortion metric is defined as
\begin{align}
d_{mse}(\pmb{x}, \pmb{y}) \triangleq \left\|\pmb{x} - \pmb{y}\right\|^{2}_2.
\end{align}
\end{definition}

\subsection{Compression-Estimation in Networks}
\label{sec:NetworkCompressionEstimationProblem}

\begin{definition}[Linear Transform Network $\mathcal{N}$]\label{def:LTN}
An~\LTN~model $\mathcal{N}$ is a communication network modeled by DAG $G = (\VV,\EDG)$, weight-set $\CC$, source/receiver nodes $(\mathcal{S}, \mathcal{T})$, sources $\mathcal{X}$, sets $\mathcal{L}_{G}$, and $\mathcal{B}_{G}$ from Definitions~\ref{def:Graph}-\ref{def:DecodingModel}. Second-order source statistics are given by $\mng{\Sigma}_{\pmb{x}}$ (Definition~\ref{def:BasicSourceModel}). The operational meaning of compression-estimation matrices in $\mathcal{L}_{G}$ and $\mathcal{B}_{G}$ is in terms of signal flows on $G$ (Definition~\ref{def:CommModel}). The desired reconstruction vectors $\{\pmb{r}_i\}_{i:v_i \in \mathcal{T}}$ have known second-order statistics $\mng{\Sigma}_{\pmb{r}_i}$ and $\mng{\Sigma}_{\pmb{r}_i \pmb{x}}$. The set $\{\pmb{\hat{r}}_i\}_{i:v_i \in \mathcal{T}}$ denotes the LLSE estimates formed at receivers (Definition~\ref{def:DecodingModel}). For noisy networks, noise variables along link $(i,j) \in \EDG$ have known covariances $\mng{\Sigma}_{\pmb{z}_{ij}}$. Power constraints are given by set $\mathcal{P}$ in Definition~\ref{def:CommModel}.
\end{definition}


Given an \LTN~graph $\mathcal{N}$, the task is to design a
\emph{network transform code}: the compression-estimation matrices in $\mathcal{L}_{G}$ and $\mathcal{B}_{G}$ to minimize the end-to-end weighted MSE distortion. Let positive weights $\{w_i\}_{i:v_i \in \mathcal{T}}$ represent the relative importance of reconstructing a signal at receiver $v_i \in \mathcal{T}$. Using indexing term $\kappa \triangleq |\VV| - |\mathcal{T}|$ for receiver nodes, we concatenate vectors $\pmb{r}_i$ as $\pmb{r} = \left[\pmb{r}_{\kappa+1};\pmb{r}_{\kappa+2};~\ldots~;\pmb{r}_{|\mathcal{V}|}\right]$ and LLSE estimates $\pmb{\hat{r}}_i$ as $\pmb{\hat{r}} = \left[\pmb{\hat{r}}_{\kappa+1}; \pmb{\hat{r}}_{\kappa+2}; ~\ldots~;\pmb{\hat{r}}_{|\mathcal{V}|}\right]$. The average weighted MSE written via a weighted $\ell^2$-norm is
\begin{align} D_{MSE,\mng{W}} & \triangleq \EE\left[\sum_{i: v_i \in \mathcal{T}} d_{mse}(\sqrt{w_i}\pmb{r}_i, \sqrt{w_i}\pmb{\hat{r}}_i)\right], \notag \\ 
& = \EE\Bigl[\bigl\|\pmb{r} - \pmb{\hat{r}}\bigl\|_{\mng{W}}^2\Bigl], \label{eqn:MSEforLTN}
\end{align}
where $\mng{W}$ contains diagonal blocks $\mng{W}_i = \sqrt{w_i}\,\mng{I}$.
\begin{remark} The distortion $D_{MSE, \mng{W}}$ is a function of the compression matrices in $\mathcal{L}_G$ and the estimation matrices in $\mathcal{B}_G$. In most network topologies, the weighted MSE distortion is non-convex over the set of feasible matrices. Even in the particular case of distributed compression~\cite{gastpar06}, currently the optimal linear transforms are not solvable in closed form.
\end{remark}

\section{Linear Signal Processing in Networks}
\label{sec:LinearProcessingNetworks}
The linear processing and filtering of source signals by an \LTN~graph $\mathcal{N}$ is modeled compactly as a linear system with inputs, outputs, and memory elements. At each time step, \LTN~nodes transmit random signals through edges/channels of the graph.

\subsection{Linear System}
\label{sec:LinearSystem}
Consider edge $(i,j) \in \mathcal{E}$ as a memory element storing random vector $\pmb{y}_{ij}$. Let $c \triangleq (\sum_{(i,j)\in \EDG} c_{ij})$ and $d \triangleq (\sum_{i:v_i\in \mathcal{T}} d_i^{-})$. The network $\mathcal{N}$ is modeled as a linear system with the following signals: (i) input sources $\{\pmb{x}_i\}_{i:v_i \in \mathcal{S}}$ concatenated as global source vector $\pmb{x} \in \mathbb{R}^{n}$; (ii) input noise variables $\{\pmb{z}_{ij}\}_{(i,j) \in \EDG}$ concatenated as global noise vector $\pmb{z} \in \mathbb{R}^{c}$; (iii) memory elements $\{\pmb{y}_{ij}\}_{(i,j) \in \EDG}$ concatenated as global state vector $\pmb{\mu}[t] \in \mathbb{R}^{c}$ at time $t$; (iv) output vectors $\{\pmb{y}_i\}_{i:v_i \in \mathcal{T}}$ concatenated as $\pmb{y} \in \mathbb{R}^{d}$.

\subsubsection{State-space Equations} The linear system\footnote{When discussing zero-delay linear coding, the time indices on vectors $\pmb{x}$, $\pmb{z}$, and $\pmb{y}_{i}$ are omitted for greater clarity of presentation.} is described by the following state-space equations for $i:v_i \in \mathcal{T}$,
\begin{align}
\pmb{\mu}[t+1] &= \mng{F}\pmb{\mu}[t] + \mng{E}\pmb{x}[t] + \mng{\tilde{E}}\pmb{z}[t], \label{eqn:LinSys1}\\
\pmb{y}_{i}[t] &= \mng{C}_i\pmb{\mu}[t] + \mng{D}_i\pmb{x}[t] + \mng{\tilde{D}}_i\pmb{z}[t]. \label{eqn:LinSys2}
\end{align}
The matrix $\mng{F} \in \mathbb{R}^{c \times c}$ is the state-evolution matrix common to all receivers, $\mng{E} \in \mathbb{R}^{c \times n}$ is the source-network connectivity matrix, and $\mng{\tilde{E}} \in \mathbb{R}^{c \times c}$ is the noise-to-network connectivity matrix. The matrices $\mng{C}_i \in \mathbb{R}^{d_i^{-} \times c}$, $\mng{D}_i \in \mathbb{R}^{d_i^{-} \times n}$, and $\mng{\tilde{D}}_i \in \mathbb{R}^{d_i^{-} \times c}$ represent how each receiver's output is related to the state, source, and noise vectors respectively. For networks considered in this paper, $\mng{D}_i = \mng{0}$ and $\mng{\tilde{D}}_i = \mng{0}$.

\subsubsection{Linear Transfer Function} A standard result in linear system theory yields the transfer function (assuming a unity indeterminate delay operator) for each receiver $v_i \in \mathcal{T}$,
\begin{align}
\pmb{y}_i & = \mng{C}_i\left(\mng{I} - \mng{F}\right)^{-1}(\mng{E}\pmb{x} + \mng{\tilde{E}}\pmb{z}), \label{eqn:LinSysEqNoise} \\
& = \mng{G}_i\pmb{x} + \mng{\tilde{G}}_i\pmb{z}, \label{eqn:LinSysEq}
\end{align}
where $\mng{G}_i \triangleq \mng{C}_i\left(\mng{I} - \mng{F}\right)^{-1}\mng{E}$ and $\mng{\tilde{G}}_i \triangleq \mng{C}_i\left(\mng{I} - \mng{F}\right)^{-1}\mng{\tilde{E}}$. For acyclic graphs, $\mng{F}$ is a nilpotent matrix and $\left(\mng{I} - \mng{F}\right)^{-1} = \mng{I} + \sum_{k=1}^{\gamma}\mng{F}^{\gamma}$ for finite integer $\gamma$. Using indexing term $\kappa$, the observation vectors collected by receivers are concatenated as $\pmb{y} = \left[\pmb{y}_{\kappa+1};~\pmb{y}_{\kappa+2};~\ldots~;~\pmb{y}_{|\mathcal{V}|}\right]$. Let
\begin{align}
\mng{T} \triangleq \left[\mng{G}_{\kappa+1};~\mng{G}_{\kappa + 2};~\ldots~;~\mng{G}_{|\mathcal{V}|}\right], \label{eqn:FullTransferFunction}
\end{align}
and let $\mng{\tilde{T}}$ be defined similarly with respect to matrices $\mng{\tilde{G}}_i$. Then the complete linear transfer function of the network $\mathcal{N}$ is $\pmb{y} = \mng{T}\pmb{x} + \mng{\tilde{T}}\pmb{z}$. Analog processing of signals without error control implies noise propagation; the additive noise $\pmb{z}$ is also linearly filtered by the network via $\mng{\tilde{T}}$.
\vspace{0.1in}
\begin{example}\label{ex:LinearSystemOfRelayNetwork} Fig.~\ref{fig:NoisyRelayNetwork} is the \LTN~graph of a noisy relay network. Let state $\pmb{\mu} = \left[\pmb{y}_{12};~\pmb{y}_{13};~\pmb{y}_{23}\right]$, $\pmb{z} = \left[\pmb{z}_{12};~\pmb{z}_{13};~\pmb{z}_{23}\right]$, and output $\pmb{y}_3 = \left[\pmb{y}_{13};~\pmb{y}_{23}\right]$. The linear system representation is given as follows,
\begin{align}
\pmb{\mu}[t+1] & = \begin{bmatrix} \mng{0} & \mng{0} & \mng{0}~ \\ \mng{0} & \mng{0} & \mng{0}~ \\ \mng{L}_{23} & \mng{0} & \mng{0}~ \end{bmatrix} \pmb{\mu}[t] + \begin{bmatrix} \,\mng{L}_{12}\, \\ \,\mng{L}_{13}\, \\ \,\mng{0}\, \end{bmatrix}\pmb{x}_1[t] + \mng{I}_{c}\pmb{z}[t], \notag \\
\pmb{y}_3[t] & = \begin{bmatrix} \;~\mng{0}~ & \,\mng{I} & \mng{0}~ \\ \;~\mng{0}~ & \,\mng{0} & \mng{I}~ \end{bmatrix} \pmb{\mu}[t]. \notag
\end{align}
By evaluating Eqn.~\eqref{eqn:LinSysEq},
\begin{align}
\pmb{y}_3[t] & = \begin{bmatrix}\;\mng{L}_{13}\; \\ \;\mng{L}_{23}\mng{L}_{12}\; \end{bmatrix}\pmb{x}_1[t] + \begin{bmatrix}~\mng{0} & \mng{I} & \mng{0}~ \\ ~\mng{L}_{23} & \mng{0} & \mng{I}~ \end{bmatrix}\pmb{z}[t].\notag
\end{align}
Dropping the time indices and writing $\pmb{x} = \pmb{x}_1$ in addition to $\pmb{y} = \pmb{y}_3$, the linear transfer function of the noisy relay network is of the following form: $\pmb{y} = \mng{T}\pmb{x} + \mng{\tilde{T}}\pmb{z}$.
\end{example}
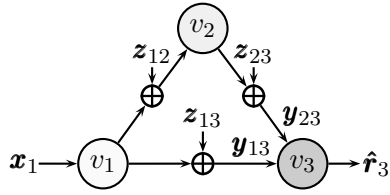
\begin{figure}[t]
\begin{center}
\psset{unit=0.45mm}
\begin{pspicture}(-11,-67)(120,-3)


\rput(-2,-31){

\pnode(7,-30){S1} \rput(3,-30){$\pmb{x}_1$}


\pscircleop[opsep=0, scale=1.232](55.5,-30){op13}
\pnode(51.5,-30){Y}
\pnode(59.5,-30){Z}
\pnode(55.5,-26){K}
\pnode(55.5,-20.5){J}
\ncline{->}{J}{op13}
\rput(55.5,-16.5){$\pmb{z}_{13}$}

\pscircleop[opsep=0, scale=1.232](40.5,-10){op12}
\pnode(37.6716,-12.8284){YP}
\pnode(43.3284,-7.1716){ZP}
\pnode(40.5,-6){KP}
\pnode(40.5,-1.5){JP}
\ncline{->}{JP}{op12}
\rput(40.5,2.5){$\pmb{z}_{12}$}

\pscircleop[opsep=0, scale=1.232](70.5,-10){op23}
\pnode(73.3284,-12.8284){YQ}
\pnode(67.6716,-7.1716){ZQ}
\pnode(70.5,-6){KQ}
\pnode(70.5,-1.5){JQ}
\ncline{->}{JQ}{op23}
\rput(70.5,2.5){$\pmb{z}_{23}$}

\cnodeput[linestyle=solid,linecolor=black,fillcolor=Sgray,fillstyle=solid](26,-30){A}{$v_1$}
\cnodeput[linestyle=solid,linecolor=black,fillcolor=Rgray,fillstyle=solid](55.5,10){H}{$v_2$}
\cnodeput[linestyle=solid,linecolor=black,fillcolor=Dgray,fillstyle=solid](85,-30){B}{$v_3$}

\rput(70,-25){$\pmb{y}_{13}$}
\rput(85,-16){$\pmb{y}_{23}$}

\pnode(102,-30){C} \ncline{->}{B}{C}
\rput(107,-30){$\pmb{\hat{r}}_3$}

\ncline{->}{A}{op13}
\ncline{->}{op13}{B}
\ncline{->}{A}{op12}
\ncline{->}{op12}{H}
\ncline{->}{H}{op23}
\ncline{->}{op23}{B}

\ncline{->}{S1}{A}


}

\end{pspicture}
\end{center} 
\caption{The \LTN~graph of a noisy relay network with $\mathcal{S} = \{v_1\}$ and $\mathcal{T} = \{v_3\}$. The linear processing of the network is modeled as a linear system with input $\pmb{x}_1$ and output $\pmb{y}_3 = [\pmb{y}_{13};~\pmb{y}_{23}]$.} \label{fig:NoisyRelayNetwork}
\end{figure}
\subsection{Layered Networks}
\begin{definition}[Layered DAG Network]\label{def:Layering} A layering of a DAG $G = (\VV, \EDG)$ is a partition of $\VV$ into disjoint subsets $\VV_1, \VV_2, \ldots, \VV_{p+1}$ such that if directed edge $(u,v) \in \EDG$, where $u \in \VV_j$ and $v \in \VV_k$, then $j > k$. A DAG layering (non-unique) is polynomial-time computable~\cite{healy02}.
\end{definition}

Given a layered partition $\{\VV_\ell\}_{\ell=1}^{p+1}$ of an \LTN~graph, source nodes $v_i \in \mathcal{S}$ with in-degree $d_{i}^{-} = 0$ may be placed in partition $\VV_{p+1}$. Similarly, receivers $v_i \in \mathcal{T}$ with out-degree $d_{i}^{+} = 0$ may be placed in partition $\VV_{1}$. The transfer function $\mng{T}$ in Eqn.~\eqref{eqn:FullTransferFunction} may be factored into a product of matrices,
\begin{align} 
\mng{T} & = \mng{T}_{1:p} \triangleq \mng{T}_1\mng{T}_2\cdot\cdot\cdot\mng{T}_p, \label{eqn:TProductForm}
\end{align}
where $\mng{T}_{\ell}$ for $1 \leq \ell \leq p$ is the linear transformation of signals between nodes in partition $\VV_{\ell+1}$ and $\VV_{\ell}$ (note the reverse ordering of the $\mng{T}_{\ell}$ with respect to the partitions $\VV_{\ell}$). If an edge exists between nodes in non-consecutive partitions, an identity transform is inserted to replicate signals between multiple layers. Due to the linearity of transforms, for any layered partition $\{\VV_{\ell}\}_{\ell = 1}^{p+1}$ of $\VV$, the layered transforms $\{\mng{T}_{\ell}\}_{\ell=1}^{p}$ can be constructed. The $\{\mng{T}_{\ell}\}_{\ell=1}^{p}$ are structured matrices comprised of sub-blocks $\mng{L}_{ij}$, identity matrices, and/or zero matrices. The block structure is determined by the network topology.
\vspace{0.1in}
\begin{example}\label{ex:LayeredEncodingTransforms} For the multiple unicast network of Fig.~\ref{fig:PosterNetwork}, a valid layered partition of $\VV$ is $\VV_1 = \{v_5, v_6\}$, $\VV_2 = \{v_4\}$, $\VV_3 = \{v_3\}$, and $\VV_{4} = \{v_1, v_2\}$. Let $\pmb{x} = [\pmb{x}_{1};~\pmb{x}_2]$, $\pmb{y} = [\pmb{y}_{5};~\pmb{y}_{6}] = [\pmb{y}_{15};~\pmb{y}_{45};~\pmb{y}_{46};~\pmb{y}_{26}]$, and let $\mng{L}_{34}$ be partitioned as $\mng{L}_{34} = \left[\mng{L}_{34}^{\prime}~~\mng{L}_{34}^{\prime\prime}\right]$. According to the layering, the transfer matrix $\mng{T}$ is factored in product form $\mng{T} = \mng{T}_1\mng{T}_2\mng{T}_3$,
\begin{equation}
\mng{T} = \begin{bmatrix} ~\mng{I} & \mng{0} & \mng{0}~ \\ ~\mng{0} & \mng{L}_{45} & \mng{0}~ \\ ~\mng{0} & \mng{L}_{46} & \mng{0}~ \\ ~\mng{0} & \mng{0} & \mng{I}~ \end{bmatrix}\begin{bmatrix} \,\mng{I} & \mng{0} & \mng{0} & \mng{0}\, \\ \,\mng{0} & \mng{L}_{34}^{\prime} & \mng{L}_{34}^{\prime\prime} & \mng{0}\, \\ \,\mng{0} & \mng{0} & \mng{0} & \mng{I}\, \end{bmatrix}\begin{bmatrix} ~\mng{L}_{15} & \mng{0}~ \\ ~\mng{L}_{13} & \mng{0}~ \\ ~\mng{0} & \mng{L}_{23}~ \\ ~\mng{0} & \mng{L}_{26}~ \end{bmatrix}. \notag
\end{equation}
\end{example}
\begin{example}\label{ex:LayeredEncodingTransformsAnotherExample} Consider the setting of Example~\ref{ex:LinearSystemOfRelayNetwork} for the relay network shown in Fig.~\ref{fig:NoisyRelayNetwork}. A valid layered partition of $\VV$ is $\VV_1 = \{v_3\}$, $\VV_2 = \{v_2\}$, $\VV_3 = \{v_1\}$. According to the layering, the transfer matrix $\mng{T}$ may be written in product form $\mng{T} = \mng{T}_1\mng{T}_2$,
\begin{align}
\mng{T} & = \begin{bmatrix} ~\mng{I} & \mng{0}~ \\ ~\mng{0} & \mng{L}_{23}~ \end{bmatrix}\begin{bmatrix} ~\mng{L}_{13}~ \\ ~\mng{L}_{12}~ \end{bmatrix}. \notag
\end{align}
\end{example}
\vspace{0.05in}
\section{Optimizing Compression-Estimation Matrices}
\label{sec:IterativeMethod}
Our optimization method proceeds iteratively over network layers. To simplify the optimization, we first assume ideal channels (high-SNR communication) for which $\pmb{y}_{ij} = \pmb{x}_{ij}$. Then the linear operation of the network $\mathcal{N}$ is $\pmb{y} = \mng{T}\pmb{x}$ with $\pmb{z} = \mng{0}$. Linear transform coding is constrained according to bandwidth compression ratios $\alpha_{ij}$.
\vspace{-0.1in}
\subsection{MSE Distortion at Receivers}
\label{sec:SumDistOfTDestNodes}
According to the linear system equations, Eqns.~\eqref{eqn:LinSys1}-\eqref{eqn:LinSysEq}, each receiver $v_i \in \mathcal{T}$ receives filtered source observations $\pmb{y}_{i} = \mng{G}_i\pmb{x}$. Receiver $v_i$ applies a linear estimator $\mng{B}_i$ to estimate signal $\pmb{r}_i$. The MSE cost of estimation is
\begin{align}
D_{i} & = \EE\Bigl[\bigl\|\pmb{r}_i -
\mng{B}_i\mng{G}_i\pmb{x}\bigl\|_{2}^2\Bigl] \notag \\
& = \TR\bigl(\mng{\Sigma}_{\pmb{r}_i}\bigl) -
2\TR\bigl(\mng{B}_i\mng{G}_i\mng{\Sigma}_{\pmb{x}\pmb{r}_i}\bigl) +
\TR\bigl(\mng{B}_i\mng{G}_i\mng{\Sigma}_{\pmb{x}}\mng{G}_i^{T}\mng{B}_i^{T}\bigl).
\label{eqn:NoiselessMSEperDest}
\end{align}
Setting the matrix derivative with respect to $\mng{B}_i$ in Eqn.~\eqref{eqn:NoiselessMSEperDest} to zero yields: $-2\mng{\Sigma}_{\pmb{r}_i \pmb{x}}\mng{G}_i^{T} +
2\mng{B}_i\mng{G}_i\mng{\Sigma}_{\pmb{x}}\mng{G}_i^{T} = 0$. For a fixed transfer function $\mng{G}_i$, the optimal LLSE matrix $\mng{B}_i^{opt}$ is
\begin{align}
\mng{B}_i^{opt} & = \mng{\Sigma}_{\pmb{r}_i \pmb{x}}\mng{G}_i^{T}
\left[\mng{G}_i \mng{\Sigma}_{\pmb{x}}\mng{G}_i^{T} \right]^{-1}.
\label{eqn:lemmaBiperDestOpt}
\end{align}
If $\mng{G}_i$ in Eqn.~\eqref{eqn:lemmaBiperDestOpt} is singular, the inverse may be replaced with a pseudo-inverse operation to compute $\mng{B}_i^{opt}$.

Let $\mng{B}$ denote a block diagonal global matrix containing individual decoding matrices $\{\mng{B}_i\}_{i:v_i \in \mathcal{T}}$ on the diagonal. For an \LTN~graph $\mathcal{N}$ with encoding transfer function $\mng{T} = \mng{T}_{1:p}$, we write the linear decoding operation of all receivers as $\pmb{\hat{r}} = \mng{B}\pmb{y}$ where $\pmb{y} = \mng{T}_{1:p}\pmb{x}$ are the observations received. The weighted MSE cost in Eqn.~\eqref{eqn:MSEforLTN} for reconstructing signals $\{\pmb{r}_i\}_{i:v_i \in \mathcal{T}}$ at all receivers is written as
\begin{align}
D_{MSE,\mng{W}} & = \EE\left[\left\|\pmb{r} -
\pmb{\hat{r}}\right\|_{\mng{W}}^2\right] \notag \\
& = \EE\left[\left\|\pmb{r} -
\mng{B}\mng{T}_{1:p}\pmb{x}\right\|_{\mng{W}}^2\right] \notag \\
& = \TR\left(\mng{W}\mng{\Sigma}_{\pmb{r}}\mng{W}^{T}\right) - 2\TR\left(\mng{W}\mng{B}\mng{T}_{1:p}\mng{\Sigma}_{\pmb{x}\pmb{r}}\mng{W}^{T}\right) \notag \\
& ~~~ +
\TR\left(\mng{W}\mng{B}\mng{T}_{1:p}\mng{\Sigma}_{\pmb{x}}\mng{T}_{1:p}^{T}\mng{B}^{T}\mng{W}^{T}\right).
\label{eqn:NoiselessMSE}
\end{align}
By construction of the weighting matrix $\mng{W}$, the MSE in Eqn.~\eqref{eqn:NoiselessMSE} is a weighted sum of individual distortions at receivers, i.e. $D_{MSE, \mng{W}} = \sum_{i:v_i \in \mathcal{T}} w_i\,D_i$.
\subsection{Computing Encoding Transforms $\mng{T}_i$}
\label{sec:ComputingEncoding}

The optimization of the network transfer function $\mng{T} = \mng{T}_{1:p}$ is more complex due to block constraints imposed by the network topology on matrices $\{\mng{T}_i\}_{i=1}^{p}$. In order to solve for a particular linear transform $\mng{T}_i$, we assume all linear transforms
$\mng{T}_{j}$, $j \neq i$ and the receivers' decoding transform $\mng{B}$ are
fixed. Then the optimal $\mng{T}_i$ is the solution to a constrained
quadratic program. To derive this, we utilize the following identities
in which $\pmb{x} = \mbox{vec}(\mng{X})$:
\begin{align}
\TR\bigl(\mng{A}^{T}\mng{X}\bigl) & = \mbox{vec}(\mng{A})^{T}\pmb{x}, 
\label{eqn:MatrixIdent1} \\
\TR\bigl(\mng{X}^{T}\mng{A}_1\mng{X}\mng{A}_2\bigl) & =
\pmb{x}^{T}(\mng{A}_2\otimes\mng{A}_1)\pmb{x}. 
\label{eqn:MatrixIdent2}
\end{align}
We write the network's linear transfer function as $\mng{T} =
\mng{T}_{1:p} = \mng{T}_{1:i-1}\mng{T}_i\mng{T}_{i+1:p}$ and
define the following matrices
\begin{align}
\mng{J}_i & \triangleq
\mng{T}_{i+1:p}\mng{\Sigma}_{\pmb{x}\pmb{r}}\mng{W}^{T}\mng{W}\mng{B}\mng{T}_{1:i-1},
\label{eqn:J1} \\
\mng{J}_i^{\prime} & \triangleq
(\mng{T}_{1:i-1})^{T}\mng{B}^{T}\mng{W}^{T}\mng{W}\mng{B}\mng{T}_{1:i-1}, \label{eqn:J2} \\
\mng{J}_i^{\prime\prime} & \triangleq
\mng{T}_{i+1:p}\mng{\Sigma}_{\pmb{x}}(\mng{T}_{i+1:p})^{T}.
\label{eqn:J3}
\end{align}
To write $D_{MSE,\mng{W}}$ in terms of the matrix variable $\mng{T}_i$,
we also define the following,
\begin{align}
p_i & \triangleq
\TR\bigl(\mng{W}\mng{\Sigma}_{\pmb{r}}\mng{W}^{T}\bigl),
\label{eqn:scalarP} \\
\pmb{p}_i & \triangleq -2\mbox{vec}\bigl(\mng{J}_i^{T}\bigl),
\label{eqn:vectorP} \\
\mng{P}_i & \triangleq \mng{J}_i^{\prime\prime} \otimes
\mng{J}_i^{\prime}, \label{eqn:matrixP}
\end{align}
where $p_i$, $\pmb{p}_i$, and $\mng{P}_i$ are a scalar, vector, and
positive semi-definite matrix respectively. The following lemma
expresses $D_{MSE, \mng{W}}$ as a function of the unknown matrix
variable $\mng{T}_i$.

\begin{lemma}\label{lemma:DistTi} Let transforms $\mng{T}_{j}$, $j \neq i$, and $\mng{B}$
be fixed. Let $\mng{J}_i$, $\mng{J}_i^{\prime}$,
$\mng{J}_i^{\prime\prime}$ be defined in
Eqns.~\eqref{eqn:J1}-\eqref{eqn:J3}, and $p_i$, $\pmb{p}_i$, and
$\mng{P}_i$ be defined in Eqns.~\eqref{eqn:scalarP}-\eqref{eqn:matrixP}. Then the weighted MSE distortion $D_{MSE, \mng{W}}$ of Eqn.~\eqref{eqn:NoiselessMSE} is a
quadratic function of $\pmb{t}_i = \mbox{vec}(\mng{T}_i)$,
\begin{align}
D_{MSE, \mng{W}} & = \pmb{t}_i^{T}\mng{P}_i\pmb{t}_i +
\pmb{p}_i^{T}\pmb{t}_i + p_i.
\label{eqn:NoiselessMSEperDestNoTraces}
\end{align}
\end{lemma}
\begin{proof} Substituting the expressions for $\mng{J}_i$, $\mng{J}_i^{\prime}$,
$\mng{J}_i^{\prime\prime}$ in Eqns.~\eqref{eqn:J1}-\eqref{eqn:J3}
into Eqn.~\eqref{eqn:NoiselessMSE} produces the intermediate equation: $D_{MSE, \mng{W}} = \TR\bigl(\mng{T}_i^{T}\mng{J}_i^{\prime} \mng{T}_i \mng{J}_i^{\prime\prime}\bigl) - 2\TR\bigl(\mng{J}_i\mng{T}_i\bigl) + p_i.$ Directly applying the vector-matrix identities of
Eqns.~\eqref{eqn:MatrixIdent1}-\eqref{eqn:MatrixIdent2} results in Eqn.~\eqref{eqn:NoiselessMSEperDestNoTraces}.
\end{proof}
\input{hybridKLT_network}
\subsection{Quadratic Program with Convex Constraints}
\label{sec:QuadraticProgramWithConvexConstraintsNoiseless}
Due to Lemma~\ref{lemma:DistTi}, the weighted MSE is a quadratic function of $\pmb{t}_i = \mbox{vec}(\mng{T}_i)$ if all other network matrices are fixed. The optimal $\mng{T}_i$ must satisfy block constraints determined by network topology. The block constraints are \emph{linear equality constraints} of the form $\mng{\Phi}_i\pmb{t}_i = \pmb{\phi}_i$. For example, if $\mng{T}_i$ contains an
identity sub-block, this is enforced by setting entries in $\pmb{t}_i$ to zero and one accordingly, via linear equality constraints.
\small
\begin{algorithm}[t]
  \caption{\textsc{Ideal-Compression-Estimation($\mathcal{N}$, $\mng{W}$, $\epsilon$)} }
  \label{alg:multipleLayerIdealNetwork}
  \center
  \begin{algorithmic}[1] \normalsize
   \STATE Identify compression matrices $\{\mng{T}_i\}_{i=1}^{p}$ and corresponding linear equalities $\{\mng{\Phi}_i, \pmb{\phi}_i\}_{i=1}^{p}$ for network $\mathcal{N}$. Identify estimation matrices $\{\mng{B}_i\}_{i:v_i \in \mathcal{T}}$. [Sec.~\ref{sec:LinearProcessingNetworks}, Sec.~\ref{sec:QuadraticProgramWithConvexConstraintsNoiseless}]
   \STATE Initialize $\{\mng{T}_{i}^{(0)}\}_{i=1}^{p}$ randomly to feasible matrices.
   \STATE Set $n = 1$, $D_{MSE, \mng{W}}(0) = \infty$.
   \REPEAT
       \STATE Compute $\{\mng{B}_{i}^{(n)}\}_{i:v_i \in \mathcal{T}}$ given $\{\mng{T}_{k}^{(n-1)}\}_{k=1}^{p}$.~[Eqn.~\eqref{eqn:lemmaBiperDestOpt}]\label{alg:stepBopt}
       \FOR{$i=1:p$}
           \STATE Compute $\mng{T}_i^{(n)}$ given $\{\mng{\Phi}_i, \pmb{\phi}_i\}$, $\{\mng{B}_{k}^{(n)}\}_{k:v_k \in \mathcal{T}}$, $\{\mng{T}_{k}^{(n)}\}_{k=1}^{(i-1)}$, $\{\mng{T}_k^{(n-1)}\}_{k=i+1}^{p}$.~[Theorem~\ref{thm:OptimalTi}]\label{alg:stepTopt}
       \ENDFOR
       \STATE Compute $D_{MSE, \mng{W}}(n)$.~[Eqn.~\eqref{eqn:NoiselessMSE}]
       \STATE Set $\Delta_{MSE,\mng{W}} = D_{MSE, \mng{W}}(n) - D_{MSE, \mng{W}}(n-1)$.
       \STATE Set $n = n + 1$.
   \UNTIL{$\Delta_{MSE,\mng{W}} \leq \epsilon$ or $n \geq N_{max}$.
   \RETURN $\{\mng{T}_{i}^{(n)}\}_{i=1}^{p}$, $\{\mng{B}_{i}^{(n)}\}_{i: v_i \in \mathcal{T}}$.
  \end{algorithmic}
\end{algorithm}} \normalsize
\vspace{0.05in}
\begin{theorem}[Optimal Encoding]\label{thm:OptimalTi}
Let encoding matrices $\mng{T}_{j}$, $j \neq i$ and decoding matrix
$\mng{B}$ be fixed. Let $\pmb{t}_i = \mbox{vec}(\mng{T}_i)$. The
optimal encoding transform $\pmb{t}_i$ is given by the following
constrained quadratic program
(QP)~\cite[Def.~4.34]{boydvandenberghe}
\begin{align}
\arg \min_{~\pmb{t}_i} & ~~~~ \pmb{t}_i^{T}\mng{P}_i\pmb{t}_i +
\pmb{p}_i^{T}\pmb{t}_i + p_i \label{eqn:QuadProg1} \\
\mbox{s. t.} & ~~~~\mng{\Phi}_i\pmb{t}_i = \pmb{\phi}_i, \notag
\end{align}
where $(\mng{\Phi}_i, \pmb{\phi}_i)$ represent linear equality
constraints on elements of $\mng{T}_i$. The solution to the above
optimization for $\pmb{t}_i$ is obtained by solving a corresponding
linear system
\begin{align}
\left[\begin{array}{cc} 2\mng{P}_i & \mng{\Phi}_i^{T} \\
\mng{\Phi}_i &
\mng{0}\end{array}\right]\left[\begin{array}{c} \pmb{t}_i \\
\pmb{\lambda} \end{array}\right] & = \left[\begin{array}{c}
-\pmb{p}_i
\\ \pmb{\phi}_i \end{array}\right]. \label{eqn:LCNLinearSystem}
\end{align}
If the constraints determined by the pair $(\mng{\Phi}_i,
\pmb{\phi}_i)$ are feasible, the linear system of
Eqn.~\eqref{eqn:LCNLinearSystem} is guaranteed to have either one or
infinitely many solutions.
\end{theorem}
\begin{proof} The QP of Eqn.~\eqref{eqn:QuadProg1} follows from
Lemma~\ref{lemma:DistTi} with additional linear equality constraints
placed on $\pmb{t}_i$. The closed form solution to the QP is derived
using Lagrange dual multipliers for the linear constraints, and the
Karush-Kuhn-Tucker (KKT) conditions. Let $f(\pmb{t}_i,
\pmb{\lambda})$ represent the Lagrangian formed with dual vector
variable $\pmb{\lambda}$ for the constraints,
\begin{align}
f(\pmb{t}_i, \pmb{\lambda}) & = \pmb{t}_i^{T}\mng{P}_{i}\pmb{t}_i + \pmb{p}_i^{T}\pmb{t}_i + p_i + \pmb{\lambda}^{T}\left(\mng{\Phi}_i\pmb{t}_i - \pmb{\phi}_i\right), \\
\nabla_{\pmb{t}_i} f(\pmb{t}_i, \pmb{\lambda}) & = 2\mng{P}_i\pmb{t}_i + \pmb{p}_i + \mng{\Phi}_i^{T}\pmb{\lambda}, \\
\nabla_{\pmb{\lambda}} f(\pmb{t}_i, \pmb{\lambda}) & =
\mng{\Phi}_i\pmb{\pmb{t}_i} - \pmb{\phi}_i.
\end{align}
Setting $\nabla_{\pmb{t}_i} f(\pmb{t}_i, \pmb{\lambda}) =
\mathbf{0}$ and $\nabla_{\pmb{\lambda}} f(\pmb{t}_i, \pmb{\lambda})
= \mathbf{0}$ yields the linear system of
Eqn.~\eqref{eqn:LCNLinearSystem}, the solutions to which are
$\pmb{t}_i$ and dual vector $\pmb{\lambda}$. Since the MSE
distortion is bounded by a minimum of zero error, the linear system
has a unique solution if $\mng{P}_i$ is full rank, or infinitely
many solutions of equivalent objective value if $\mng{P}_i$ is
singular.
\end{proof}
\begin{remark} Beyond linear constraints, several other convex constraints on matrix variables could be applied within the quadratic program. For example, the $\ell_1$-norm of a vector $\pmb{x} \in \mathbb{R}^{n}$ defined by $\|\pmb{x}\|_{1} \triangleq \sum_i |x_i|$ is often used in compressed sensing to enforce \emph{sparsity}.
\end{remark}

\subsection{An Iterative Algorithm}
Algorithm~\ref{alg:multipleLayerIdealNetwork} defines an iterative
method to optimize all encoding matrices $\{\mng{T}_i\}_{i=1}^{p}$ and the global decoding
matrix $\mng{B}$ for an \LTN~graph. The iterative algorithm begins with the random initialization of the encoding matrices $\{\mng{T}_i\}_{i=1}^{p}$ subject to size specifications and linear equality constraints given by $\{\mng{\Phi}_i\}_{i=1}^{p}$ and $\{\pmb{\phi}_i\}_{i=1}^{p}$. The iterative method proceeds by solving for the optimal $\mng{B}$ transform first. Similarly, with $\mng{T}_{j}, j \neq i$ and $\mng{B}$ fixed, the optimal $\mng{T}_i$ is computed using Theorem~\ref{thm:OptimalTi}. The iterative method proceeds for $n \leq N_{max}$ iterations or until the difference in error $\Delta_{MSE, \mng{W}}$ is less than a prescribed tolerance $\epsilon$.

\subsection{Convergence to Stationary Points}
A key property of Algorithm~\ref{alg:multipleLayerIdealNetwork} is
the convergence to a stationary point (either local minimum or
saddle-point) of the weighted MSE.
\begin{theorem}[Local Convergence]\label{thm:ConvergenceAlg1}
Denote the network's linear transfer function after the $n$-th outer-loop iteration
in Algorithm~\ref{alg:multipleLayerIdealNetwork} by $\mng{T}^{(n)}$, and the block-diagonal global decoding transform by $\mng{B}^{(n)}$ which contains matrices $\{\mng{B}_i^{(n)}\}_{i:v_i \in \mathcal{T}}$ on the diagonal. Let $\pmb{\hat{r}}^{(n)} = \mng{B}^{(n)}\mng{T}^{(n)}\pmb{x}$ denote the estimate of desired signal $\pmb{r}$. Then
\begin{align}
\EE\left[\left\|\pmb{r} -
\pmb{\hat{r}}^{(n)}\right\|^2_{\mng{W}}\right] \geq
\EE\left[\left\|\pmb{r} -
\pmb{\hat{r}}^{(n+1)}\right\|^2_{\mng{W}}\right],
\end{align}
i.e., the weighted MSE distortion is a nonincreasing function of the
iteration number $n$.
\end{theorem}
\begin{proof} In Step~\ref{alg:stepBopt} of
Algorithm~\ref{alg:multipleLayerIdealNetwork}, with matrices
$\{\mng{T}_k^{(n-1)}\}_{k=1}^{p}$ fixed, the optimal transform
$\mng{B}^{(n)}$ is determined to minimize $D_{MSE, \mng{W}}$. The
current transform $\mng{B}^{(n-1)}$ is feasible within the
optimization space which implies that the MSE distortion cannot increase. In
Step~\ref{alg:stepTopt} of the inner loop, with matrices $\mng{B}^{(n)}$, $\{\mng{T}_{k}^{(n)}\}_{k=1}^{(i-1)}$, and $\{\mng{T}_{k}^{(n-1)}\}_{k=i+1}^{p}$ fixed, Theorem~\ref{thm:OptimalTi}
computes the optimal transform $\mng{T}_i^{(n)}$ to minimize
$D_{MSE, \mng{W}}$. A similar argument shows that the error term
cannot increase. The distortion sequence $\{D_{MSE, \mng{W}}(n)\}$
is nonincreasing and nonnegative; hence $\lim_{n \rightarrow
\infty} D_{MSE,\mng{W}}(n) = \inf\{D_{MSE,\mng{W}}(n)\}$ by monotone
convergence.
\end{proof}
\begin{table}[t]
\center \caption{\small \textsc{A ``Hybrid'' Linear Transform Network}}
\label{tbl:HybridNetwork}
\newcolumntype{K}{>{\raggedright\arraybackslash$}X<{$}}
\begin{tabularx}{.485\textwidth}{*{2}{K}}\toprule[1.2pt] \small \textbf{Network~Modes} & Bandwidth~ \\\midrule[0.7pt]
Distributed & c \leq \lfloor\frac{c_{34}}{2}\rfloor \\
Hybrid & \lceil\frac{c_{34}}{2}\rceil < c < c_{34} \\
Point~to~Point & c_{34} \leq c \\ \bottomrule[1.2pt]
\end{tabularx}
\end{table}
\begin{remark} The local convergence in Theorem~\ref{thm:ConvergenceAlg1} is affected by several factors: (i) The covariance structure $\mng{\Sigma}_{\pmb{x}}$ of the source; (ii) The DAG structure of $G$; (iii) The schedule of iterative optimization of local matrices and factorization of $\mng{T}$ into the $\mng{T}_i$; (iv) The random initialization of $\{\mng{T}_i\}_{i=1}^{p}$. In practice, multiple executions of Algorithm~\ref{alg:multipleLayerIdealNetwork} increase the probability of converging to a global minimum.
\end{remark}

\subsection{Example: A Multi-Hop Network}
Consider the noiseless multi-hop network of Fig.~\ref{fig:HybridKLTFig} in which a relay aggregates, compresses and/or forwards its observations to a receiver. The network is a hybrid combination of a distributed and point-to-point network.

\begin{example}[``Hybrid Network'']\label{ex:HybridNetwork} High-dimensional, correlated
signals $\pmb{x}_1 \in \mathbb{R}^{n_1}$ and $\pmb{x}_2 \in \mathbb{R}^{n_2}$ are observed at nodes $v_1$ and $v_2$ where $n_1 = n_2 = 15$ dimensions. The covariance $\mng{\Sigma}_{\pmb{x}}$ of the global source $\pmb{x} = [\pmb{x}_1;~\pmb{x}_2]$ was generated as follows for the experiment, ensuring $\mng{\Sigma}_{\pmb{x}} \succ \mng{0}$. The diagonal entries $(i,i)$ of $\mng{\Sigma}_{\pmb{x}}$ were selected as $15 + 2U_{ii}$, and off-diagonal entries $(i,j)$ for $j > i$ were selected as $1 + 2U_{ij}$ where $U_{ii}$ and $U_{ij}$ are $i.i.d.$ uniform random variables over the interval $[0,1]$.

The linear transfer function is factored in the form $\mng{T} = \mng{T}_1 \mng{T}_2$ where $\mng{T}_1 = \mng{L}_{34}$ and
\begin{align}
\mng{T}_2 & = \left[\begin{array}{cc} \mng{L}_{13} & \mng{0} \\
\mng{0} & \mng{L}_{23} \end{array}\right]. \notag
\end{align}
The target reconstruction at $v_4$ is the entire signal $\pmb{r}_4 = \pmb{x}$. The bandwidth $c_{34} = 11$,
while bandwidth $c = c_{13} = c_{23}$ is varied for the experiment. Depending on the
amount of bandwidth $c$, the network operates in one of the modes given in
Table~\ref{tbl:HybridNetwork}. Fig.~\ref{fig:HybridKLTFig}(b) plots
the sum distortion vs. compression performance, and
Fig.~\ref{fig:HybridKLTFig}(c) plots the convergence of
Algorithm~\ref{alg:multipleLayerIdealNetwork} for the operating
point $c = 6$, $c_{34} = 11$.
\end{example}

\section{Noisy Networks}
\label{sec:NoisyModel}

We now analyze communication for networks with non-ideal channels: $\pmb{y}_{ij} = \pmb{x}_{ij} + \pmb{z}_{ij}$. Edges $(i,j)$ represent vector Gaussian channels. Network communication is limited according to both bandwidth compression ratios $\alpha_{ij}$ and signal-to-noise ratios $SNR_{ij}$. We simplify optimization of subspaces by restricting attention to single-layer multi-source, multi-receiver networks for which $\VV = \mathcal{S} \cup \mathcal{T}$. In this case, the linear transfer function is $\pmb{y} = \mng{T}\pmb{x} + \pmb{z}$, i.e. the noise is additive but not filtered over multiple network layers.

\subsection{MSE Distortion at Receivers} \label{sec:SumDistOfTDestNodesWithNoise}

Each receiver $v_i \in \mathcal{T}$ receives observations $\pmb{y}_i = \mng{G}_i\pmb{x} + \pmb{z}_i$ where $\pmb{z}_i$ is the noise to $v_i$. The MSE distortion for reconstructing $\pmb{r}_i$ at receiver $v_i$ is given by,
\begin{align}
\tilde{D}_{i} & = \TR\bigl(\mng{\Sigma}_{\pmb{r}}\bigl) -
2\TR\bigl(\mng{B}_i\mng{G}_i\mng{\Sigma}_{\pmb{x}\pmb{r}_i}\bigl) +
\TR\bigl(\mng{B}_i\mng{\Sigma}_{\pmb{z}_i}\mng{B}_i^{T}\bigl) \notag \\
& ~~~+ \TR\bigl(\mng{B}_i\mng{G}_i\mng{\Sigma}_{\pmb{x}}\mng{G}_i^{T}\mng{B}_i^{T}\bigl).
\label{eqn:NoisyMSEperDest}
\end{align}
Setting the matrix derivative with respect to $\mng{B}_i$ in Eqn.~\eqref{eqn:NoisyMSEperDest} to zero yields the optimal linear transform $\mng{B}_i$ (cf. Eqn.~\eqref{eqn:lemmaBiperDestOpt}),
\begin{align}
\mng{B}_i^{opt} & = \mng{\Sigma}_{\pmb{r}_i \pmb{x}}\mng{G}_i^{T}
\Big[\mng{G}_i \mng{\Sigma}_{\pmb{x}}\mng{G}_i^{T} +
\mng{\Sigma}_{\pmb{z}_i} \Big]^{-1}.
\label{eqn:lemmaBiperDestOptNoisy}
\end{align}
Combining the LLSE estimates as $\pmb{\hat{r}} = \mng{B}\pmb{y}$, where $\pmb{y} = \mng{T}\pmb{x} + \pmb{z}$, the weighted MSE for all receivers is given by
\begin{align}
\tilde{D}_{MSE,\mng{W}} & = \EE\Bigl[\bigl\|\pmb{r} -
\pmb{\hat{r}}\bigl\|_{\mng{W}}^2\Bigl] \notag \\
& = \EE\Bigl[\bigl\|\pmb{r} - \mng{B}(\mng{T}\pmb{x} + \pmb{z})\bigl\|^2_{\mng{W}}\Bigl] \notag \\
& =
\TR\bigl(\mng{W}\mng{B}\mng{T}\mng{\Sigma}_{\pmb{x}}\mng{T}^{T}\mng{B}^{T}\mng{W}^{T}\bigl) - 2\TR\bigl(\mng{W}\mng{B}\mng{T}\mng{\Sigma}_{\pmb{x}\pmb{r}}\mng{W}^{T}\bigl)  \notag \\
& ~~~+ \TR\bigl(\mng{W}\mng{\Sigma}_{\pmb{r}}\mng{W}^{T}\bigl) +
\TR\bigl(\mng{W}\mng{B}\mng{\Sigma}_{\pmb{z}}\mng{B}^{T}\mng{W}^{T}\bigl).
\label{eqn:NoisyMSE}
\end{align}
By construction of the weighting matrix $\mng{W}$, the MSE in Eqn.~\eqref{eqn:NoisyMSE} is a weighted sum of individual distortions at receivers, i.e. $\tilde{D}_{MSE,\mng{W}} = \sum_i w_i\,\tilde{D}_i$.

\subsection{Computing Encoding Transform $\mng{T}$}
\label{sec:ComputingEncodingNoisy}

For noisy networks, power constraints on channel inputs limit the amount of amplification of transmitted signals. For single-layer networks, let $v_i \in \mathcal{S}$ be a source node with observed signal $\pmb{x}_i$. A power constraint on the input to channel $(i,j) \in \EDG$ is given by
\begin{align}
\EE[\|\pmb{x}_{ij}\|^2_2] = \EE[\|\mng{L}_{ij}\pmb{x}_{i}\|^2_2] = \TR\bigl(\mng{L}_{ij}\mng{\Sigma}_{\pmb{x}_i}\mng{L}_{ij}^{T}\bigl)
\leq P_{ij}. \label{eqn:PowerConstraintIsQuad}
\end{align}
The power constraint in Eqn.~\eqref{eqn:PowerConstraintIsQuad} is a
quadratic function of the entries of the global linear transform
$\mng{T}$. More precisely, let $\pmb{\ell}_{ij} =
\mbox{vec}(\mng{L}_{ij})$ and $\pmb{t} = \mbox{vec}(\mng{T})$. Since
$\pmb{t}$ contains all variables of $\pmb{\ell}_{ij}$, we may write
$\pmb{\ell}_{ij} = \mng{J}_{ij}\pmb{t}$ where $\mng{J}_{ij}$ selects
variables from $\pmb{t}$. Using the matrix-vector identities of
Eqn.~\eqref{eqn:MatrixIdent2}, the power constraint in
Eqn.~\eqref{eqn:PowerConstraintIsQuad} can be written as
\begin{align}
\TR\left(\mng{L}_{ij}\mng{\Sigma}_{\pmb{x}_i}\mng{L}_{ij}^{T}\right) & = \pmb{\ell}_{ij}^{T}\left(\mng{\Sigma}_{\pmb{x}_i} \otimes \mng{I}\right)\pmb{\ell}_{ij} \notag \\ & = \pmb{t}^{T}\mng{J}_{ij}^{T}\left(\mng{\Sigma}_{\pmb{x}_i} \otimes \mng{I}\right)\mng{J}_{ij}\pmb{t}. \label{eqn:QuadPowerConstraintDerivation}
\end{align}
Letting $\mng{\Gamma}_{ij} \triangleq\mng{J}_{ij}^{T}\left(\mng{\Sigma}_{\pmb{x}_i} \otimes \mng{I}\right)\mng{J}_{ij}$, the quadratic constraint is $\pmb{t}^{T}\mng{\Gamma}_{ij}\pmb{t} \leq P_{ij}$. The matrix $\mng{\Gamma}_{ij}$ is a symmetric, positive semi-definite matrix. Thus a power constraint is a quadratic, convex constraint.


\small
\begin{algorithm}[t]
  \caption{\textsc{Noisy-Compression-Estimation($\mathcal{N}, \mng{W}, \epsilon$)}}
  \label{alg:singlelayerNoisyNetwork}
  \center
  \begin{algorithmic}[1] \normalsize
   \STATE Identify compression matrix $\mng{T}$ and corresponding linear equality constraints $(\mng{\Phi}, \pmb{\phi})$, and quadratic power constraints $\{(\mng{\Gamma}_{ij}, P_{ij})\}_{(i,j) \in \EDG}$. Identify estimation matrices $\{\mng{B}_i\}_{i:v_i \in \mathcal{T}}$.~[Sec.~\ref{sec:LinearProcessingNetworks}, Sec.~\ref{sec:ComputingEncodingNoisy}]
   \STATE Initialize $\mng{T}^{(0)}$ randomly to a feasible matrix.
   \STATE Set $n = 1$, $\tilde{D}_{MSE, \mng{W}}(0) = \infty$.
   \REPEAT
       \STATE Compute $\{\mng{B}_{i}^{(n)}\}_{i:v_i \in \mathcal{T}}$ given $\mng{T}^{(n-1)}$.~[Eqn.~\eqref{eqn:lemmaBiperDestOptNoisy}] \label{alg:stepBoptNoisy}
       \STATE Compute $\mng{T}^{(n)}$ given $\{\mng{B}_{i}^{(n)}\}_{i:v_i \in \mathcal{T}}$,~$(\mng{\Phi}, \pmb{\phi})$, $\{(\mng{\Gamma}_{ij}, P_{ij})\}_{(i,j) \in \EDG}$.~[Theorem~\ref{thm:OptimalT1NoisePower}]\label{alg:stepToptNoisy}
       \STATE Compute $\tilde{D}_{MSE, \mng{W}}(n)$.~[Eqn.~\eqref{eqn:NoisyMSE}]
       \STATE Set $\tilde{\Delta}_{MSE,\mng{W}} = \tilde{D}_{MSE,\mng{W}}(n) - \tilde{D}_{MSE,\mng{W}}(n-1)$.
       \STATE Set $n = n + 1$.
   \UNTIL{$\tilde{\Delta}_{MSE,\mng{W}} \leq \epsilon$ or $n \geq N_{max}$.
   \RETURN $\mng{T}^{(n)}$ and $\{\mng{B}_{i}^{(n)}\}_{i:v_i \in \mathcal{T}}$.
  \end{algorithmic}
\end{algorithm}} \normalsize

\subsection{Quadratic Program with Convex Constraints}

As in Section~\ref{sec:ComputingEncoding}, we use the vector form
$\pmb{t} = \mbox{vec}(\mng{T})$ to enforce linear equality constraints
$\mng{\Phi}\pmb{t} = \pmb{\phi}$. For noisy networks, we include power constraints $\pmb{t}^{T}\mng{\Gamma}_{ij}\pmb{t} \leq P_{ij}$ for each channel $(i,j) \in \EDG$. For a fixed global decoding transform $\mng{B}$, the distortion $\tilde{D}_{MSE, \mng{W}}$ of Eqn.~\eqref{eqn:NoisyMSE}
is again a quadratic function of $\pmb{t}$. Using the compact notation
\begin{align}
p & \triangleq
\TR\bigl(\mng{W}\mng{\Sigma}_{\pmb{r}}\mng{W}^{T}\bigl) +
\TR\bigl(\mng{W}\mng{B}\mng{\Sigma}_{\pmb{z}}\mng{B}^{T}\mng{W}^{T}\bigl), \label{eqn:scalarPNoise} \\
\pmb{p} & \triangleq
-2\mbox{vec}\bigl(\mng{B}^{T}\mng{W}^{T}\mng{W}\mng{\Sigma}_{\pmb{r}\pmb{x}}\bigl),
\label{eqn:vectorPNoise} \\
\mng{P} & \triangleq \mng{\Sigma}_{\pmb{x}} \otimes
\mng{B}^{T}\mng{W}^{T}\mng{W}\mng{B}, \label{eqn:matrixPNoise}
\end{align}
a derivation identical to that of Lemma~\ref{lemma:DistTi} yields
$\tilde{D}_{MSE, \mng{W}} = \pmb{t}^{T}\mng{P}\pmb{t} +
\pmb{p}^{T}\pmb{t} + p$. The optimal encoding transform $\mng{T}$
for single-layer noisy networks is solvable via a quadratic program with quadratic constraints (QCQP), following the development of Eqns.~\eqref{eqn:scalarPNoise}-\eqref{eqn:matrixPNoise}, and the power constraints given in Eqns.~\eqref{eqn:PowerConstraintIsQuad}-\eqref{eqn:QuadPowerConstraintDerivation}; cf.~Theorem~\ref{thm:OptimalTi}.
\begin{theorem}[Optimal Encoding $\mng{T}$ for Noisy \LTN]\label{thm:OptimalT1NoisePower}
Let $\mathcal{N}$ be a single-layer \LTN, $\mng{B}$ be the fixed
decoding transform, and $\pmb{t} = \mbox{vec}(\mng{T})$ be the
encoding transform. The optimal encoding $\pmb{t}$ is the solution
to the following quadratic program with quadratic constraints
(QCQP):
\begin{align}
\arg \min_{~\pmb{t}} & ~~~~ \pmb{t}^{T}\mng{P}\pmb{t} +
\pmb{p}^{T}\pmb{t} + p \label{eqn:QuadProg1Noisy} \\
\mbox{s. t.} & ~~~~\mng{\Phi}\pmb{t} = \pmb{\phi}, \notag \\
~~ & ~~~~\pmb{t}^{T}\mng{\Gamma}_{ij}\pmb{t} \leq P_{ij}, ~~~(i,j) \in \EDG, \notag
\end{align}
where $(\mng{\Phi}, \pmb{\phi})$ represent linear equality constraints (dictated by network topology), and $\{(\mng{\Gamma}_{ij}, P_{ij})\}_{(i,j) \in \EDG}$ represent quadratic power constraints on variables of $\mng{T}$.
\end{theorem}
\begin{remark} A quadratic program with linear and convex quadratic constraints is solvable efficiently via standard convex program solvers; the time complexity depends polynomially on the number of matrix variables and constraints.
\end{remark}
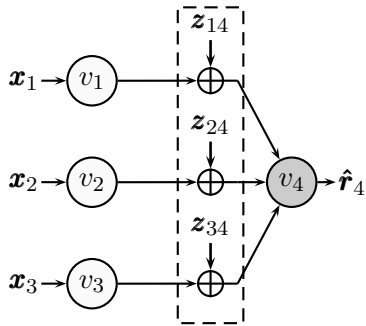
\begin{figure}[t]
\begin{center}
\psset{unit=0.45mm}
\begin{pspicture}(-11,-105)(115,-3)

\rput(0, 20){





\rput(0,-55){

\pnode(11,0){S1} \pnode(11,-30){S2} \pnode(11,-60){S3}
\rput(6,0){$\pmb{x}_{1}$} \rput(6,-30){$\pmb{x}_{2}$}
\rput(6,-60){$\pmb{x}_{3}$}

\psframe[linecolor=black,linestyle=dashed](51,22)(71, -72)

\cnodeput[linestyle=solid,linecolor=black,fillcolor=Sgray,fillstyle=solid](26,0){A}{$v_1$}
\cnodeput[linestyle=solid,linecolor=black,fillcolor=Sgray,fillstyle=solid](26,-30){B}{$v_2$}
\cnodeput[linestyle=solid,linecolor=black,fillcolor=Sgray,fillstyle=solid](26,-60){F}{$v_3$}
\cnodeput[linestyle=solid,linecolor=black,fillcolor=Dgray,fillstyle=solid](85,-30){E}{$v_4$}
\pnode(69, 0){C} \pnode(69, -30){D} \pnode(69, -60){R}

\pnode(57,0){X} \pnode(57,-30){Y} \pnode(57,-60){Z}

\pnode(98,-30){G} \ncline{->}{E}{G}
\rput(103,-30){$\pmb{\hat{r}}_4$}

\pscircle[linecolor=black](61,0){4}
\pscircle[linecolor=black](61,-30){4}
\pscircle[linecolor=black](61,-60){4}
\psline[linecolor=black](61,4)(61,-4)
\psline[linecolor=black](61,-26)(61,-34)
\psline[linecolor=black](61,-56)(61,-64)

\psline[linecolor=black, linewidth=1pt]{->}(61,-18)(61,-26)
\psline[linecolor=black, linewidth=1pt]{->}(61,12)(61,4)
\psline[linecolor=black, linewidth=1pt]{->}(61,-48)(61,-56)
\rput(61,-13){$\pmb{z}_{24}$} \rput(61,17){$\pmb{z}_{14}$}
\rput(61,-43){$\pmb{z}_{34}$}

\ncline{->}{A}{X} \ncline{->}{B}{Y} \ncline{->}{F}{Z}
\ncline{-}{X}{C} \ncline{-}{Y}{D} \ncline{-}{Z}{R}

\ncline{->}{C}{E} \ncline{->}{D}{E} \ncline{->}{R}{E}

\ncline{->}{S1}{A} \ncline{->}{S2}{B} \ncline{->}{S3}{F}


}




}

\end{pspicture}
\end{center}
\vspace{0.1in} \caption{A block diagram of a distributed,
noise/power limited \LTN. Each source node transmits signal
projections of a vector $\pmb{x}_i \in \mathbb{R}^{4}$ to a decoder over a vector arbitrary white Gaussian noise (AWGN) channel.} \vspace{-0.1in}
\label{fig:DistributedKLTFig}
\end{figure}
\subsection{Iterative Algorithm and Convergence}

Algorithm~\ref{alg:singlelayerNoisyNetwork} defines an iterative algorithm for single-layer, noise/power limited networks. In addition to subspace selection, the amount of power per subspace is determined iteratively. The iterative method alternates between optimizing the global decoding transform $\mng{B}$ and the global encoding transform $\mng{T}$, ensuring that network topology and power constraints are satisfied. As in Theorem~\ref{thm:ConvergenceAlg1}, the weighted MSE distortion is a nonincreasing function of the iteration number, i.e. $\tilde{D}_{MSE,\mng{W}}(n) \geq \tilde{D}_{MSE,\mng{W}}(n+1)$. While convergence to a stationary point is guaranteed, the
optimization space is highly complex-- a globally optimal solution is not guaranteed.

\subsection{Example: A Distributed Noisy Network}
Fig.~\ref{fig:DistributedKLTFig} diagrams a classic example of a
distributed network with multiple source (sensor) nodes transmitting
signal projections to a central decoder. Each source node is power
constrained and must transmit a compressed description of its
observed signal over a noisy vector channel.
\input{distributedKLT_spread_second_part_graphs}
\begin{example}[Distributed \LTN]\label{ex:DistributedKLTNoisePower} In Fig.~\ref{fig:DistributedKLTFig}, the global source $\pmb{x} = [\pmb{x}_1;~\pmb{x}_2;~\pmb{x}_3]$ is chosen to be a \emph{jointly Gaussian} vector with $n
= 12$ dimensions, and $n_i = 4$ for each of $|\mathcal{S}| = 3$
source nodes. Here, we specify the exact distribution of $\pmb{x}$ in order to provide information-theoretic lower bounds. We set the covariance of $\pmb{x}$ to be Gauss-Markov with $\rho = 0.8$,
\begin{align}
\mng{\Sigma}_{\pmb{x}} & = \left[\begin{array}{ccccc} 1 & \rho &
\rho^2 & \ldots & \rho^{11} \\ \rho & 1 & \rho & \ldots & \rho^{10}
\\ \rho^2 & \rho & 1 & \ldots & \rho^{9} \\ \vdots & \vdots & \vdots & \ddots & \vdots
\\ \rho^{11} & \rho^{10} & \rho^{9} & \ldots & 1\end{array}\right]. \notag
\end{align}
The network structure is specified by bandwidths $c_{14} =
c_{24} = c_{34} = c$. The global encoding transform $\mng{T}$ is
block-diagonal with matrices $\mng{L}_{14}$, $\mng{L}_{24}$, and
$\mng{L}_{34}$ on the diagonal. The compression ratio is varied
equally for each source node, $\alpha = \frac{c}{n_i}$ where $n_i =
4$. The noise variables $\pmb{z}_{ij}$ are $i.i.d.$ Gaussian
random vectors with zero-mean and identity covariances. The power constraints are
set as $P_1 = P_2 = P_3 = c(SNR)$, where $SNR_{ij} = SNR$ for all
links. The goal of destination $v_4$ is to reconstruct the entire
source $\pmb{r}_4 = \pmb{x}$. Fig.~\ref{fig:DistributedKLTSecondPartGraphsFig}(a) plots the performance of \LTN~optimization for varying $\alpha$ and $SNR$ ratios as well as cut-set lower bounds for linear coding based on convex relaxations. Cut-set lower bounds for linear coding for this example are explained further in Section~\ref{sec:NoisyDistributedKLTCutSet}. Fig.~\ref{fig:DistributedKLTSecondPartGraphsFig}(b) plots cut-set bounds based on information theory which are explained further in Sections~\ref{sec:AppendixCutSetBoundsInfoTheory} and~\ref{sec:NoisyDistributedKLTCutSetInfoTheoryBound}.
\end{example}
\begin{remark}[Comparison with~\cite{gastpar06, giannakis07}] For this example, as the $SNR \rightarrow \infty$, the error $\tilde{D}_{MSE}$ approaches the error associated to the distributed KLT~\cite{gastpar06} where channel noise was not considered. In~\cite{giannakis07}, the authors model the effects of channel noise; however, they do not provide cut-set lower bounds. In addition, the iterative optimization of the present paper optimizes all compression matrices simultaneously per iteration and allows arbitrary convex constraints, as opposed to the schemes in both~\cite{gastpar06, giannakis07} which optimize the encoding matrix of each user separately per iteration.
\end{remark}



\section{Cut-Set Lower Bounds}
\label{sec:CutsetBounds}

In this section, we derive lower bounds on the minimum MSE distortion possible for linear compression and estimation of correlated signals in the \LTN~model. Our main technique is to relax an arbitrary acyclic graph along all possible graph cuts to point-to-point networks with side information. The cut-set bounds provide a performance benchmark for the iterative methods of Sections~\ref{sec:IterativeMethod}-\ref{sec:NoisyModel}. 

\subsection{Point-to-Point Network with Side Information}
\label{sec:RelaxNetwork}
\begin{figure}[t]
\begin{center}
\psset{unit=0.45mm}
\begin{pspicture}(-11,-59)(115,-3)


\rput(-2,-10){

\pnode(7,-30){S1} \rput(3,-30){$\pmb{x}$}

\psframe[linecolor=black,linestyle=dashed](16,-19)(36, -41)
\psframe[linecolor=black,linestyle=dashed](75,-19)(95, -41)

\pscircle[linecolor=black](55.5,-30){4}
\psline[linecolor=black](55.5,-26)(55.5,-34)
\psline[linecolor=black](51.5,-30)(59.5,-30)
\pnode(51.5,-30){Y}
\pnode(59.5,-30){Z}
\pnode(55.5,-26){K}
\pnode(55.5,-12){J}
\ncline{->}{J}{K}
\rput(55.5,-8){$\pmb{z}$}

\cnodeput[linestyle=solid,linecolor=black,fillcolor=Sgray,fillstyle=solid](26,-30){A}{$v_1$}
\cnodeput[linestyle=solid,linecolor=black,fillcolor=Dgray,fillstyle=solid](85,-30){B}{$v_2$}

\pnode(65,-55){D} \pnode(85, -55){E} \pnode(46,-5){F}
\pnode(26,-5){G} \ncline{-}{D}{E} 
\ncline{->}{E}{B}

\rput(60,-55){$\pmb{s}$} 

\pnode(102,-30){C} \ncline{->}{B}{C}
\rput(107,-30){$\pmb{\hat{r}}$}

\ncline{->}{A}{Y}
\ncline{->}{Z}{B}

\ncline{->}{S1}{A}


}

\end{pspicture}
\end{center}
\vspace{0.1in} \caption{A point-to-point network with side information
$\pmb{s}$ at the receiver. In the case of additive noise $\pmb{z}$, the input to the channel is
power-constrained so that $E[\|\pmb{x}_{12}\|_2^{2}] \leq P$. }\vspace{-0.1in} \label{fig:CutsetBoundDiagram}
\end{figure}
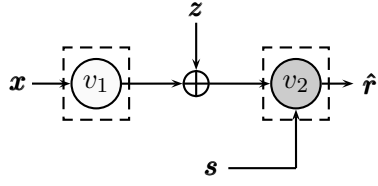

Consider the point-to-point network of Fig.~\ref{fig:CutsetBoundDiagram}. Source node $v_1$ compresses source $\pmb{x} \in \mathbb{R}^{n}$ via a linear transform $\mng{L}_{12}$. The signal $\pmb{x}_{12} \in \mathbb{R}^{c_{12}}$ is transmitted where $\pmb{x}_{12} = \mng{L}_{12}\pmb{x}$ and $\EE[\| \pmb{x}_{12}\|_2^{2}] \leq P$. Receiver $v_2$ computes a linear estimate of desired signal $\pmb{r} \in \mathbb{R}^{r}$ using observations $\pmb{y}_{12} = \pmb{x}_{12} + \pmb{z}$ and side information $\pmb{s} \in \mathbb{R}^{s}$ as follows,
\begin{align}
\pmb{\hat{r}} & = \mng{B}\begin{bmatrix} \pmb{y}_{12} \\ \pmb{s} \end{bmatrix} = \begin{bmatrix} \mng{B}_{11} & \mng{B}_{12} \end{bmatrix}\begin{bmatrix} \pmb{y}_{12} \\ \pmb{s} \end{bmatrix}.
\end{align}
The decoding transform $\mng{B}$ is here partitioned into two sub-matrices $\mng{B}_{11}$ and $\mng{B}_{12}$. We will find it convenient to define the following random vectors,
\begin{align}
\pmb{\xi} & \triangleq \pmb{x} - \mng{\Sigma}_{\pmb{x}\pmb{s}}\mng{\Sigma}_{\pmb{s}}^{-1}\pmb{s},
\label{eqn:XIDef} \\
\pmb{\nu} & \triangleq \pmb{r} - \mng{\Sigma}_{\pmb{r}\pmb{s}}\mng{\Sigma}_{\pmb{s}}^{-1}\pmb{s}.
\label{eqn:NUDef}
\end{align}
Signals $\pmb{\xi}$ and $\pmb{\nu}$ are innovation vectors. For example, $\pmb{\xi}$ is the difference between $\pmb{x}$ and the linear least squares estimate of $\pmb{x}$ given $\pmb{s}$ which is equivalent to $\mng{\Sigma}_{\pmb{x}\pmb{s}}\mng{\Sigma}_{\pmb{s}}^{-1}\pmb{s}$.

\subsection{Case I: Ideal Vector Channel}

In the ideal case, $P = \infty$ or $\pmb{z} = 0$. The weighted, linear minimum MSE distortion of the point-to-point network with side information is obtained by solving
\begin{align}
D_{ideal}^{*} & = \min_{~\mng{L}_{12}, \mng{B}} ~~~~~~
\EE\Bigl[\bigl\|\pmb{r} - \pmb{\hat{r}}\bigl\|^2_{\mng{W}}\Bigl], \notag \\
& = \min_{~\mng{L}_{12}, \mng{B}_{11}, \mng{B}_{12}} ~ \EE\Bigl[\bigl\|\pmb{r} - (\mng{B}_{11}\mng{L}_{12}\pmb{x} + \mng{B}_{12}\pmb{s})\bigl\|^2_{\mng{W}}\Bigl].
\label{eqn:OptLowerBound1}
\end{align}
The following theorem specifies the solution to Eqn.~\eqref{eqn:OptLowerBound1}.
\vspace{0.05in}
\begin{theorem}[Ideal Network Relaxation]\label{thm:ElementaryNetworkSolution}
Let $\pmb{x} \in \mathbb{R}^{n}$, $\pmb{s} \in \mathbb{R}^{s}$, and $\pmb{r} \in \mathbb{R}^{r}$ be zero-mean random vectors with given full-rank covariance matrices $\mng{\Sigma}_{\pmb{x}}$, $\mng{\Sigma}_{\pmb{s}}$, $\mng{\Sigma}_{\pmb{r}}$ and cross-covariances $\mng{\Sigma}_{\pmb{rx}}$, $\mng{\Sigma}_{\pmb{rs}}$, $\mng{\Sigma}_{\pmb{xs}}$. Let $\pmb{\xi}$ and $\pmb{\nu}$ be the innovations defined in Eqn~\eqref{eqn:XIDef} and Eqn.~\eqref{eqn:NUDef} respectively. The solution to the minimization of Eqn.~\eqref{eqn:OptLowerBound1} over matrices $\mng{L}_{12} \in \mathbb{R}^{c_{12} \times n}$, $\mng{B}_{11} \in \mathbb{R}^{r \times c_{12}}$, and $\mng{B}_{12} \in \mathbb{R}^{r \times s}$ is obtained in closed form as
\begin{align}
D_{ideal}^{*} & = \TR\bigl(\mng{\Sigma}_{\pmb{\nu}}\mng{W}^{T}\mng{W}\bigl) - \sum_{j=1}^{c_{12}} \lambda_{j},
\label{eqn:MMSEforElementaryNetwork}
\end{align}
where $\{\lambda_j\}_{j=1}^{c_{12}}$ are the $c_{12}$ largest eigenvalues of the matrix
$\mng{W}\mng{\Sigma}_{\pmb{\nu}\pmb{\xi}}\mng{\Sigma}_{\pmb{\xi}}^{-1}\mng{\Sigma}_{\pmb{\xi}\pmb{\nu}}\mng{W}^{T}$.
\end{theorem}
\begin{proof} The optimization in Eqn.~\eqref{eqn:OptLowerBound1} is simplified by first determining the LMMSE optimal $\mng{B}_{12}$ transform in terms of $\mng{B}_{11}$ and $\mng{L}_{12}$: $\mng{B}_{12}^{opt} = \mng{\Sigma}_{\pmb{r}\pmb{s}}\mng{\Sigma}_{\pmb{s}}^{-1} - \mng{B}_{11}\mng{L}_{12}\mng{\Sigma}_{\pmb{x}\pmb{s}}\mng{\Sigma}_{\pmb{s}}^{-1}$.
Plugging $\mng{B}_{12}^{opt}$ into Eqn.~\eqref{eqn:OptLowerBound1} yields a minimization over
$\mng{B}_{11}$ and $\mng{L}_{12}$ only. By grouping and rearranging variables in terms of innovation vectors $\pmb{\xi}$ and $\pmb{\nu}$,
\begin{align}
D_{ideal}^{*} & = \min_{~\mng{L}_{12}, \mng{B}_{11}}
~\EE\Bigl[\bigl\|\pmb{\nu} -
\mng{B}_{11}\mng{L}_{12}\pmb{\xi}\bigl\|^2_{\mng{W}}\Bigl].
\label{eqn:OptLowerBound3}
\end{align}
The optimization of Eqn.~\eqref{eqn:OptLowerBound3} is that of an
equivalent point-to-point network with input signal $\pmb{\xi}$ and
desired reconstruction $\pmb{\nu}$, without side information. Eqn.~\eqref{eqn:OptLowerBound3} is in standard form and solvable using canonical correlation analysis as detailed
in~\cite[p.~368]{brillinger81}. The optimal value $D_{ideal}^{*}$ is given in Eqn.~\eqref{eqn:MMSEforElementaryNetwork} in terms of the eigenvalues of $\mng{W}\mng{\Sigma}_{\pmb{\nu}\pmb{\xi}}\mng{\Sigma}_{\pmb{\xi}}^{-1}\mng{\Sigma}_{\pmb{\xi}\pmb{\nu}}\mng{W}^{T}$.
\end{proof}

\subsection{Case II: Additive Noise and Power Constraints}

In the case of additive noise $\pmb{z}$ (here with assumed covariance $\mng{\Sigma}_{\pmb{z}} = \mng{I}$ for compactness) and a power-constrained input to the vector channel, the weighted, linear minimum MSE distortion is obtained by solving
\begin{align}
D_{noisy}^{*} & = \!\!\min_{~\mng{L}_{12}, \mng{B}_{11}, \mng{B}_{12}} \EE\Bigl[\bigl\|\pmb{r} - (\mng{B}_{11}(\mng{L}_{12}\pmb{x} + \pmb{z}) + \mng{B}_{12}\pmb{s})\bigl\|^2_{\mng{W}}\Bigl], \notag \\
& ~~~~~~ \mbox{s.t.}~~~ \TR[\mng{L}_{12}\mng{\Sigma}_{\pmb{x}}\mng{L}_{12}^{T}] \leq P.
\label{eqn:OptLowerBoundWithNoise0}
\end{align}
Again, by solving for the optimal LMMSE matrix $\mng{B}_{12}$ and grouping terms in the resulting optimization according to innovation vectors $\pmb{\xi}$ and $\pmb{\nu}$,
\begin{align}
D_{noisy}^{*} & = \min_{~\mng{L}_{12}, \mng{B}_{11}} ~~
\EE\Bigl[\bigl\|\pmb{\nu} - (\mng{B}_{11}(\mng{L}_{12}\pmb{\xi} + \pmb{z}))\bigl\|^2_{\mng{W}}\Bigl], \notag \\
& ~~~~~~ \mbox{s.t.}~~~ \TR[\mng{L}_{12}\mng{\Sigma}_{\pmb{x}}\mng{L}_{12}^{T}] \leq P.
\label{eqn:OptLowerBoundWithNoise}
\end{align}

\begin{remark}
The exact solution to Eqn.~\eqref{eqn:OptLowerBoundWithNoise} involves handling a quadratic power constraint and a rank constraint due to the reduced-dimensionality of $\mng{L}_{12}$. In~\cite[Theorem~4]{giannakis07}, a related optimization problem was solved via a Lagrangian relaxation. For our problem, we take a simpler approach using a semi-definite programming (SDP) relaxation. We first note that $D_{noisy}^{*} \geq D_{ideal}^{*}$. In the high-SNR regime, the two distortion values are asymptotically equivalent. Therefore, we compute a good approximation for the distortion $D_{noisy}^{*}$ in the low-SNR regime via the following SDP relaxation.
\end{remark}
\vspace{0.15in}
\begin{theorem}[SDP Relaxation]\label{thm:SDPRelaxation}
Consider random vectors $\pmb{x}$, $\pmb{s}$, $\pmb{r}$, $\pmb{\xi}$, $\pmb{\nu}$, and matrices $\mng{L}_{12}$, $\mng{B}_{11}$ as defined in Theorem~\ref{thm:ElementaryNetworkSolution}. In addition, let random vector $\pmb{z}$ have zero-mean and covariance $\mng{\Sigma}_{\pmb{z}} = \mng{I}$. Let $\mng{\Psi} \triangleq \mng{L}_{12}^{T}\mng{L}_{12}$ and $\mng{\Phi} \in \mathbb{R}^{r \times r}$ be an arbitrary positive semi-definite matrix where $r$ is the dimension of random vector $\pmb{r}$. The following lower bound applies,
\begin{align}
D_{noisy}^{*} & \geq \min_{~\mng{\Phi}, \mng{\Psi}} ~~
\TR[\mng{\Phi}] + \TR\bigl[\mng{W}\bigl[\mng{\Sigma}_{\pmb{\nu}} - \mng{\Sigma}_{\pmb{\nu}\pmb{\xi}}\mng{\Sigma}_{\pmb{\xi}}^{-1}\mng{\Sigma}_{\pmb{\xi}\pmb{\nu}}\bigl]\mng{W}^{T}\bigl], \notag \\
& ~~~~~~ \mbox{s.t.}~~~ \TR[\mng{\Sigma}_{\pmb{x}}\mng{\Psi}] \leq P, ~~\mng{\Psi} \succeq \mng{0}, \notag \\
& ~~~~~~ \mbox{~~~}~~~ \left[\begin{array}{cc} \mng{\Phi} & \mng{W}\mng{\Sigma}_{\pmb{\nu}\pmb{\xi}}\mng{\Sigma}_{\pmb{\xi}}^{-1} \\ \mng{\Sigma}_{\pmb{\xi}}^{-1}\mng{\Sigma}_{\pmb{\xi}\pmb{\nu}}\mng{W}^{T} & \mng{\Sigma}_{\pmb{\xi}}^{-1} + \mng{\Psi}\end{array}\right] \succeq \mng{0}.
\label{eqn:SDPRelaxOne}
\end{align}
\end{theorem}
The proof of Theorem~\ref{thm:SDPRelaxation} is based on a rank relaxation as detailed in the Appendix. The power constraint is still enforced in Eqn.~\eqref{eqn:SDPRelaxOne}. In the low-SNR regime, power allocation over subspaces dominates the error performance. If we denote the solution to the SDP of Theorem~\ref{thm:SDPRelaxation} as $D_{sdp}^{*}$, we arrive at the following characterization,
\begin{align}
D_{noisy}^{*} \geq \max\{D_{ideal}^{*}, D_{sdp}^{*}\}. \label{eqn:BoundingNaveen}
\end{align}



\subsection{Cut-Set Lower Bounds for Linear Coding}
\label{sec:GeneralCutSetBound}

Consider an \LTN~graph $\mathcal{N}$ with source nodes $\mathcal{S}
\subset \VV$ and receivers $\mathcal{T} \subset \VV$. We assume
that $\mathcal{S} \cap \mathcal{T} = \emptyset$, i.e. the set of
sources and receivers are disjoint. The total bandwidth and total power across a
\emph{cut} $\mathcal{F} \subset \VV$ are defined respectively as
\begin{align}
C(\mathcal{F}) & = \sum_{\begin{subarray}{c}
        jk \in \mathcal{E} \\ j \in \mathcal{F},~k \in \mathcal{F}^{c}
      \end{subarray}}
 c_{jk}, \\
 P(\mathcal{F}) & = \sum_{\begin{subarray}{c}
        jk \in \mathcal{E} \\ j \in \mathcal{F},~k \in \mathcal{F}^{c}
      \end{subarray}}
 P_{jk},
\end{align}
where the edge set $\mathcal{E}$ and bandwidths $c_{jk}$
were defined in Section~\ref{sec:ProblemStatement}. The edges of the
graph are directed, hence the bandwidth across a cut accounts for the
$c_{ij}$ only for those edges directed from node $v_i$ to $v_j$. In the following theorem, the notation $\pmb{x}_{\mathcal{F}}$ denotes the concatenation of vectors $\pmb{x}_i : v_i \in \mathcal{F}$. The set $\mathcal{F}^{c}$ denotes the complement of $\mathcal{F}$ in $\mathcal{\VV}$.
\begin{definition}
$D_{ideal}^{*}\left[\pmb{x}, \pmb{r} \bigl| \pmb{s}; c, \mng{W}\right]$ represents the distortion $D_{ideal}^{*}$ computed with the weighted norm via $\mng{W}$ for the ideal point-to-point network with input $\pmb{x}$, bandwidth $c$, reconstruction vector $\pmb{r}$, and side information to receiver $\pmb{s}$. Similarly, $D_{noisy}^{*}\left[\pmb{x}, \pmb{r} \bigl| \pmb{s}; c, P, \mng{W}\right]$ represents the distortion $D_{noisy}^{*}$ for a noisy point-to-point network with channel-input power constraint $P$ and noise vector $\pmb{z}$ with zero-mean with identity covariance.
\end{definition}
\begin{theorem}[Cut-Set Lower Bounds]\label{thm:CutsetLowerBoundsLast}
Let $\mathcal{N}$ be an arbitrary \LTN~graph with source nodes
$\mathcal{S}$ and receivers $\mathcal{T}$. Let $\mathcal{F} \subset \VV$ be
a cut of the graph. For ideal channel communication,
\begin{align} \EE\Bigl[\bigl\|\pmb{r}_{\mathcal{F}^{c}} -
\pmb{\hat{r}}_{\mathcal{F}^{c}}\bigl\|^2_{\mng{W}}\Bigl] & \geq
D_{ideal}^{*}\Big[\pmb{x}_{\mathcal{F}},
\pmb{r}_{\mathcal{F}^{c}}\Big| \pmb{x}_{\mathcal{F}^{c}}; C(\mathcal{F}), \mng{W} \Big].
\end{align}
In the case of noisy channel communication over network $\mathcal{N}$ with additive channel noise $\pmb{z}_{ij}$ (assumed zero-mean, identity covariance),
\begin{align}
\EE\Bigl[& \bigl\|\pmb{r}_{\mathcal{F}^{c}} -
\pmb{\hat{r}}_{\mathcal{F}^{c}}\bigl\|^2_{\mng{W}}\Bigl] \notag \\
& \geq D_{noisy}^{*}\Big[\pmb{x}_{\mathcal{F}},
\pmb{r}_{\mathcal{F}^{c}}\Big| \pmb{x}_{\mathcal{F}^{c}}; C(\mathcal{F}), P(\mathcal{F}), \mng{W} \Big].
\end{align}
\end{theorem}
\begin{proof} The \LTN~graph is partitioned into two sets $\mathcal{F}$ and $\mathcal{F}^{c}$. The source nodes $v_i \in \mathcal{F}$ are merged as one source
``super'' node, and the receivers $v_i \in \mathcal{F}^{c}$ are
merged into one receiver ``super'' node. The maximum bandwidth and maximum power between the source and receiver are $C(\mathcal{F})$ and $P(\mathcal{F})$ respectively. The random vector $\pmb{x}_{\mathcal{F}^{c}}$ represents those signals with channels to the receiver super node, not accounted for in the cut $\mathcal{F}$; hence, this information is given as side information (a relaxation) to the receiver. The relaxed network after the merging process is the point-to-point network of Fig.~\ref{fig:CutsetBoundDiagram} with noise $\pmb{z}$ of dimension equal to the bandwidth $C(\mathcal{F})$ of the cut, and provides a lower bound on the MSE distortion $\EE\Bigl[\bigl\|\pmb{r}_{\mathcal{F}^{c}} -
\pmb{\hat{r}}_{\mathcal{F}^{c}}\bigl\|^2_{\mng{W}}\Bigl]$ at receivers $v_i \in \mathcal{F}^{c}$.
\end{proof}
\begin{remark}
The total number of distinct cuts $\mathcal{F}$ separating sources and receivers is
$(2^{|\mathcal{S}|}-1)(2^{|\mathcal{T}|}-1)$. For a particular cut, there exists a continuum of lower bounds for multi-receiver networks depending on the choice of weighting $\mng{W}$.
\end{remark}

\subsection{Example: Cut-Set Lower Bounds for Linear Coding}
\label{sec:NoisyDistributedKLTCutSet}

In Fig.~\ref{fig:DistributedKLTSecondPartGraphsFig}(a), cut-set lower bounds for linear coding are illustrated based on Theorem~\ref{thm:CutsetLowerBoundsLast} for a distributed noisy network. The bounds are depicted for the cut that separates all sources from the receiver. Due to our approximation method in Eqn.~\eqref{eqn:BoundingNaveen} based on the SDP relaxation, the lower bounds show tight agreement in the low-SNR and high-SNR asymptotic regimes.

\input{last_multipleunicast_example}
\subsection{Cut-Set Lower Bound From Information Theory}
\label{sec:AppendixCutSetBoundsInfoTheory}

For the point-to-point communication scenario illustrated in Fig.~\ref{fig:CutsetBoundDiagram}, the information-theoretically optimal performance can be determined precisely. Consider an $\ell$-length sequence $\{(\pmb{x}[t], \pmb{s}[t])\}_{t=1}^{\ell}$ of jointly $i.i.d.$ random vectors. The source node $v_1$ has access to the source sequence $\{\pmb{x}[t]\}_{t=1}^{\ell}$. We will assume throughout that $\pmb{r}$ (respectively $\pmb{r}[t]$) is a deterministic function of $(\pmb{x}, \pmb{s})$ (respectively $(\pmb{x}[t], \pmb{s}[t])$). The goal of receiver $v_2$ is to minimize the \emph{average} MSE distortion $D_{\ell} = \EE\left[\frac{1}{\ell}\sum_{t=1}^{\ell} \|\pmb{r}[t] - \pmb{\hat{r}}[t]\|_2^{2}\right]$ where the reconstruction sequence $\{\pmb{\hat{r}}[t]\}_{t=1}^{\ell}$ is generated based on access to side information $\{\pmb{s}[t]\}_{t=1}^{\ell}$ and the sequence of channel output vectors. We study the performance in the limit as $\ell \rightarrow \infty$ and denote $D \triangleq D_{\infty}$.

\subsubsection{Source-Channel Separation} We establish a lower bound by combining the data processing inequality with the definitions of Wyner-Ziv rate-distortion function and channel capacity. Specifically, by straightforward extension of~\cite{wz76}, the minimum rate $R(D)$ required to reconstruct $\{\pmb{r}[t]\}_{t=1}^{\infty}$ at distortion $D$ is given by $R(D) = \min I(\pmb{x}; \pmb{u}|\pmb{s})$ where the minimization is over all ``auxiliary'' random vectors $\pmb{u}$ for which $p(\pmb{u}, \pmb{x}, \pmb{s}) = p(\pmb{u}|\pmb{x})p(\pmb{x}, \pmb{s})$ and for which $\EE[\|\pmb{r} - \EE[\pmb{r}|\pmb{u},\pmb{s}]\|_2^{2}] \leq D$. Furthermore, by definition of the channel capacity $C(P)$ between $v_1$ and $v_2$, $C(P) = \max_{p(\pmb{x}_{12}): \EE\left[\|\pmb{x}_{12}\|_2^{2}\right] \leq P} I(\pmb{x}_{12}; \pmb{y}_{12})$.\footnote{The notation in information theory vs. signal processing differs. The term $I(\pmb{x}_{12}; \pmb{y}_{12})$ denotes the mutual information between random vectors whereas the term $p(\pmb{x}_{12})$ indicates a probability distribution.} Source-channel separation applies to the scenario of Fig.~\ref{fig:CutsetBoundDiagram}, and in a nearly identical proof as detailed in~\cite[Thm.~1.10]{gastparphd},
\begin{align}
R(D) \leq C(P). \label{eqn:SourceChannelSep}
\end{align}

\subsubsection{$R(D)$ for Jointly Gaussian Sources} If $\{(\pmb{r}[t], \pmb{x}[t], \pmb{s}[t])\}$ form an $i.i.d.$ sequence of jointly Gaussian random vectors, then $R(D)$ is equal to the conditional rate-distortion function~\cite[Appendix~II]{gastpar06},
\begin{align}
R_{c}(D) = \min_{p(\pmb{\hat{r}}|\pmb{x}, \pmb{s}): \EE\left[\|\pmb{r} - \pmb{\hat{r}}\|_2^{2}\right] \leq D} I(\pmb{x};\pmb{\hat{r}}|\pmb{s}). \label{eqn:ConditionalRofD}
\end{align}

\subsubsection{Capacity of the Vector AWGN Channel} If the channel noise $\pmb{z}$ is a Gaussian random vector with zero mean and covariance $\mng{\Sigma}_{\pmb{z}} = \mng{I}$, the capacity of the channel in Fig.~\ref{fig:CutsetBoundDiagram} with bandwidth $c_{12}$ and power constraint $P$ is
\begin{align}
C(P) & = \frac{c_{12}}{2}\log_2 \left[1 + \frac{P}{c_{12}}\right]. \label{eqn:CapacityPtoP}
\end{align}

\subsubsection{Cut-set Bound} We utilize Eqn.~\eqref{eqn:SourceChannelSep} to obtain an information-theoretic lower bound to the distortion achievable in any network of the type considered in this paper. An arbitrary graph is reduced via graph cuts to point-to-point networks. The following theorem collects the known information-theoretic results discussed. 
\begin{theorem}[Cut-Set Bounds: Info. Theory]\label{thm:InfoTheoryCutsetLowerBoundsLast}
Let $\mathcal{N}$ be an arbitrary \LTN~graph with vector AWGN channels. Consider a cut $\mathcal{F} \subset \VV$ separating the graph into a point-to-point network with bandwidth $C(\mathcal{F})$ and power $P(\mathcal{F})$. Let $R(D^{*}_{opt})$ be the rate-distortion function for the source $\pmb{x}_{\mathcal{F}}$ with side information $\pmb{x}_{\mathcal{F}^{c}}$ and reconstruction $\pmb{r}_{\mathcal{F}^{c}}$.\footnote{We assume that $\pmb{r}_{\mathcal{F}^{c}}$ is a deterministic function of the global source $\pmb{x}$.} Then
\begin{align}
R(D^{*}_{opt}) & \leq \frac{C(\mathcal{F})}{2}\log_2 \left[1 + \frac{P(\mathcal{F})}{C(\mathcal{F})}\right]. \label{eqn:InfoTheoryLowerBound}
\end{align}
\end{theorem}


\subsection{Example: Cut-Set Lower Bound From Information Theory}
\label{sec:NoisyDistributedKLTCutSetInfoTheoryBound}

For the noisy network in Example~\ref{ex:DistributedKLTNoisePower}, consider cut $\mathcal{F} = \{v_1, v_2, v_3\}$. The source signal $\pmb{x}_{\mathcal{F}} = \pmb{x} = [\pmb{x}_1; \pmb{x}_2; \pmb{x}_3]$ is jointly Gaussian, the side information is absent, and $\pmb{r}_{\mathcal{F}^{c}} = \pmb{x}$. Denote the eigenvalues of the source $\pmb{x}_{\mathcal{F}}$ as $\{\lambda_{\pmb{x},i}\}_{i=1}^{n}$. Evaluating Eqn.~\eqref{eqn:ConditionalRofD} as in~\cite[Appendix~II]{gastpar06}, optimal source coding corresponds to reverse water-filling over the eigenvalues (see also~\cite[Chap.~10]{coverthomas}),
\vspace{-0.05in}
\begin{align}
R_c(D^{*}_{opt}) & = \sum_{i=1}^{n} \max \left\{\frac{1}{2} \log_2 \frac{\lambda_{\pmb{x},i}}{D_i}, 0\right\}, \notag
\end{align}
\vspace{-0.05in}
\begin{subnumcases}{\!\!\!\!\!\!\!\!\!\!\!\!\!\!\!\mbox{where}~D_{i} = }
\theta & $\mbox{if}~\theta < \lambda_{\pmb{x}, i}$ \notag \\
\lambda_{\pmb{x},i} & $\mbox{if}~\theta \geq \lambda_{\pmb{x}, i}$ \notag
\end{subnumcases}
and where $\theta$ is chosen such that $\sum_{i=1}^{n} D_i = D^{*}_{opt}$. The lower bound of Eqn.~\eqref{eqn:InfoTheoryLowerBound} is plotted in Fig.~\ref{fig:DistributedKLTSecondPartGraphsFig}(b) for two different bandwidth compression ratios.

\subsection{Example: Multi-Source, Multi-Receiver Network} \label{sec:ResultsAndSimulations}

\begin{example}[Multiple Unicast]\label{ex:MultipleUnicast} In Fig.~\ref{fig:ResultsMultipleUnicast}, the global source $\pmb{x} = [\pmb{x}_1;~\pmb{x}_2]$ where $\pmb{x}_1 \in \mathbb{R}^{4}$ and $\pmb{x}_2 \in \mathbb{R}^{4}$. The correlation structure of $\pmb{x}$ is given by the following matrices,
\begin{equation}
\begin{pmat}[{|}]
\mng{\Sigma}_{11} & \mng{\Sigma}_{12} \cr\- \mng{\Sigma}_{21} &
\mng{\Sigma}_{22} \cr
\end{pmat} = \begin{pmat}[{...|...}] 2.4 & 1.1 & 0.4 & 0.0 & 0.1 & 0.1 & 0.0 & 0.1 \cr
1.1 & 1.7 & 0.8 & 0.4 & 0.0 & 0.2 & 0.2 & 0.1 \cr 0.4 & 0.8 & 1.2 &
0.0 & 0.2 & 0.6 & 0.1 & 0.3 \cr 0.0 & 0.4 & 0.0 & 0.8 & 0.3 & 0.0 &
0.1 & 0.0 \cr\- 0.1 & 0.0 & 0.2 & 0.3 & 1.1 & 0.1 & 0.2 & 0.0 \cr
0.1 & 0.2 & 0.6 & 0.0 & 0.1 & 1.2 & 0.2 & 0.1 \cr 0.0 & 0.2 & 0.1 &
0.1 & 0.2 & 0.2 & 1.0 & 0.6 \cr 0.1 & 0.1 & 0.3 & 0.0 & 0.0 & 0.1 &
0.6 & 1.2 \cr
\end{pmat}. \label{eqn:CovarianceMatrixLastExample}
\end{equation}
The network structure is specified by bandwidths $c_{ij}$ as labeled in Fig.~\ref{fig:ResultsMultipleUnicast}(a). The factorization of the global linear transform $\mng{T}$ was given in Example~\ref{ex:LayeredEncodingTransforms} of Section~\ref{sec:IterativeMethod}.

The distortion region for the network in the case when node $v_5$ estimates $\pmb{r}_5 = \pmb{x}_1$, and node $v_6$ estimates $\pmb{r}_6 = \pmb{x}_2$ is given in Fig.~\ref{fig:ResultsMultipleUnicast}(b). A direct link exists from each source to receiver. However, if the desired reconstruction at the receivers is switched as in Fig.~\ref{fig:ResultsMultipleUnicast}(c), the channel from $v_3$ to $v_4$ must be shared fully and becomes a bottleneck. The cut-set bounds of interest are shown in dotted lines. The shaded region depicts the points achievable via the iterative method of Section~\ref{sec:IterativeMethod}. In Fig.~\ref{fig:ResultsMultipleUnicast}(c), the upper and lower bounds are not tight everywhere--even if one receiver is completely ignored, the resulting problem is still a distributed compression problem for which tight bounds are not known. The achievable curve was generated by taking the convex hull of $32$ points corresponding to weighting ratios $\frac{w_5}{w_6} \in [\frac{1}{100}, 100]$.

In Table~\ref{tbl:ComparisonOfTransforms}, we compare the results of
linear transform design methods for the minimum sum distortion point (weighting ratio $\frac{w_5}{w_6} = 1$).
\begin{itemize} 
\item \emph{Random Projections}-- Each entry for all compression matrices is selected from the standard normal distribution. The sum distortion $D_5 + D_6$ is averaged over $10^2$ random compression matrices selected for all nodes.
\item \emph{Routing and Network Coding (Ad-Hoc)}-- For the scenario in Fig.~\ref{fig:ResultsMultipleUnicast}(b), nodes $v_1$ and $v_2$ project their signal onto the principal eigenvectors of $\mng{\Sigma}_{11}$ and $\mng{\Sigma}_{22}$ respectively. Routing permits each receiver to receive the best two eigenvector projections from its corresponding source, as well as an extra projection from the other source. For Fig.~\ref{fig:ResultsMultipleUnicast}(c), using a simple ``network coding'' strategy of adding signals at $v_3$, one receiver is able to receive its best two eigenvector projections, but the other receiver can only receive one best eigenvector projection. \item \emph{Iterative~QP~Optimization}-- Linear transforms are designed using the iterative method of
Section~\ref{sec:IterativeMethod}.
\item \emph{Lower Bound}-- The minimum sum distortion possible due to the cut-set lower bound of Theorem~\ref{thm:CutsetLowerBoundsLast}.
\end{itemize}
\end{example}

\begin{table}[t]
\center \caption{\small \textsc{Comparison of Reduced-Dimension Linear Transforms}}
\label{tbl:ComparisonOfTransforms}
\newcolumntype{K}{>{\raggedright\arraybackslash$}X<{$}}
\begin{tabularx}{.485\textwidth}{*{5}{K}}\toprule[1.2pt] \scriptsize & & & Fig.~5(b) & Fig.~5(c) \\ \textbf{Design Method} & & & D_{5} + D_{6} & D_{5} + D_{6} \\\midrule[0.7pt]
Random~Projections & & & 4.3170 & 6.3471 \\ 
Routing~and~Network~Coding & & & 2.7029 & 3.8170 \\
Iterative~QP~Optimization & & & 2.3258 & 2.6165 \\
\left<Lower~Bound\right> & & & 2.3243 & 2.3243 \\\bottomrule[1.2pt]
\end{tabularx}
\end{table}

\section{Conclusion}
\label{sec:Conclusion}

The linear transform network (\LTN) was proposed to model the aggregation, compression, and estimation of correlated random signals in directed, acyclic graphs. For both noiseless and noisy \LTN~graphs, a new iterative algorithm was introduced for the joint optimization of reduced-dimension network matrices. Cut-set lower bounds were introduced for zero-delay linear coding based on convex relaxations. Cut-set lower bounds for optimal coding were introduced based on information-theoretic principles. The compression-estimation tradeoffs were analyzed for several example networks. A future challenge remains to compute tighter lower bounds and relaxations for non-convex network optimization problems. Reduced-dimension linear transforms have potential applications in data fusion and sensor networks. The idea of exploiting correlations between network signals to reduce data transmission, and the idea of approximate reconstruction as opposed to exact recovery at receivers may lead to further advances in networking.

\appendix

\section{Proof of Theorem~\ref{thm:SDPRelaxation}}
\label{sec:Appendix}

Starting from the optimization in Eqn.~\eqref{eqn:OptLowerBoundWithNoise}, the LLSE optimal matrix $\mng{B}_{11}^{opt} = \mng{\Sigma}_{\pmb{\nu}\pmb{\xi}}\mng{L}_{12}^{T}(\mng{L}_{12}\mng{\Sigma}_{\pmb{\xi}}\mng{L}_{12}^{T} + \mng{I})^{-1}$, assuming $\mng{\Sigma}_{\pmb{z}} = \mng{I}$. Substituting this expression and simplifying the objective function in Eqn.~\eqref{eqn:OptLowerBoundWithNoise},
\begin{align}
D_{noisy}^{*} & = \min_{~\mng{L}_{12}} ~~
\TR\left[\mng{W}\mng{\Sigma}_{\pmb{\nu}}\mng{W}^{T}\right] \notag \\
& ~~~ + \TR\left[\mng{W}\mng{\Sigma}_{\pmb{\nu}\pmb{\xi}}\mng{L}_{12}^{T}\left[\mng{L}_{12}\mng{\Sigma}_{\pmb{\xi}}\mng{L}_{12}^{T}+ \mng{I}\right]^{-1}\mng{L}_{12}\mng{\Sigma}_{\pmb{\xi}\pmb{\nu}}\mng{W}^{T}\right] \notag \\
& ~~~~~~ \mbox{s.t.}~~~ \TR\left[\mng{L}_{12}\mng{\Sigma}_{\pmb{x}}\mng{L}_{12}^{T}\right] \leq P.
\label{eqn:OptLowerBoundWithNoiseN100}
\end{align}
Applying the Woodbury (matrix-inversion) identity~\cite[C.4.3]{boydvandenberghe} to the objective function and simplifying terms,
\begin{align}
D_{noisy}^{*} & = \min_{~\mng{L}_{12}} ~~
\TR\left[\mng{W}\mng{\Sigma}_{\pmb{\nu}}\mng{W}^{T}\right] - \TR\left[\mng{W}\mng{\Sigma}_{\pmb{\nu}\pmb{\xi}}\mng{\Sigma}_{\pmb{\xi}}^{-1}\mng{\Sigma}_{\pmb{\xi}\pmb{\nu}}\mng{W}^{T}\right] \notag \\
& ~~ + \TR\left[\mng{W}\mng{\Sigma}_{\pmb{\nu}\pmb{\xi}}\mng{\Sigma}_{\pmb{\xi}}^{-1}\left[\mng{\Sigma}_{\pmb{\xi}}^{-1}+ \mng{L}_{12}^{T}\mng{L}_{12}\right]^{-1}\mng{\Sigma}_{\pmb{\xi}}^{-1}\mng{\Sigma}_{\pmb{\xi}\pmb{\nu}}\mng{W}^{T}\right] \notag \\
& ~~~~~~ \mbox{s.t.}~~~ \TR\left[\mng{L}_{12}\mng{\Sigma}_{\pmb{x}}\mng{L}_{12}^{T}\right] \leq P.
\label{eqn:OptLowerBoundWithNoiseN200}
\end{align}
Introducing a positive semi-definite matrix $\mng{\Phi}$ such that $\mng{\Phi} \succeq \mng{W}\mng{\Sigma}_{\pmb{\nu}\pmb{\xi}}\mng{\Sigma}_{\pmb{\xi}}^{-1}\bigl[\mng{\Sigma}_{\pmb{\xi}}^{-1}+ \mng{L}_{12}^{T}\mng{L}_{12}\bigl]^{-1}\mng{\Sigma}_{\pmb{\xi}}^{-1}\mng{\Sigma}_{\pmb{\xi}\pmb{\nu}}\mng{W}^{T}$, written equivalently in Schur-complement form~\cite[A.5.5]{boydvandenberghe}, and setting $\mng{\Psi} = \mng{L}_{12}^{T}\mng{L}_{12} \in \mathbb{R}^{n \times n}$ as a rank $c_{12}$ matrix,
\begin{align}
D_{noisy}^{*} & = \min_{~\mng{\Phi}, \mng{\Psi}} ~~
\TR\left[\mng{\Phi}\right] + \TR\left[\mng{W}\bigl[\mng{\Sigma}_{\pmb{\nu}} - \mng{\Sigma}_{\pmb{\nu}\pmb{\xi}}\mng{\Sigma}_{\pmb{\xi}}^{-1}\mng{\Sigma}_{\pmb{\xi}\pmb{\nu}}\bigl]\mng{W}^{T}\right], \notag \\
& ~~~~~~ \mbox{s.t.}~~~ \TR\left[\mng{\Sigma}_{\pmb{x}}\mng{\Psi}\right] \leq P, ~~\mng{\Psi} \succeq \mng{0}, ~~\mbox{rank}\left[\mng{\Psi}\right] = c_{12}, \notag \\
& ~~~~~~ \mbox{~~~}~~~ \left[\begin{array}{cc} \mng{\Phi} & \mng{W}\mng{\Sigma}_{\pmb{\nu}\pmb{\xi}}\mng{\Sigma}_{\pmb{\xi}}^{-1} \\ \mng{\Sigma}_{\pmb{\xi}}^{-1}\mng{\Sigma}_{\pmb{\xi}\pmb{\nu}}\mng{W}^{T} & \mng{\Sigma}_{\pmb{\xi}}^{-1} + \mng{\Psi}\end{array}\right] \succeq \mng{0}.
\label{eqn:SDPRelaxN1000}
\end{align}
Dropping the rank constraint yields the relaxation of Eqn.~\eqref{eqn:SDPRelaxOne}.

\appendices


\section*{Acknowledgment}
The authors would like to thank the anonymous reviewers for their valuable comments and suggestions.


\ifCLASSOPTIONcaptionsoff
  \newpage
\fi



%



\bibliographystyle{ieeetr}
\bibliography{naveenBIB}

\end{document}